\documentclass[11pt]{article}

\usepackage{fullpage}
\usepackage{setspace}
\usepackage{parskip}
\usepackage{titlesec}
\usepackage[section]{placeins}
\usepackage{xcolor}
\usepackage{breakcites}
\usepackage{lineno}
\usepackage{hyphenat}
\usepackage{rotating}

\usepackage[margin=1in]{geometry}
\usepackage{appendix}
\usepackage{bm,subfigure,enumitem}

\def\vb{{\bm{b}}}

\def\vf{{\bm{f}}}

\def\vh{{\bm{h}}}

\def\vr{{\bm{r}}}
\def\vs{{\bm{s}}}

\def\vv{{\bm{v}}}
\def\vw{{\bm{w}}}

\def\vy{{\bm{y}}}
\def\vz{{\bm{z}}}

\def\mA{{\bm{A}}}
\def\mB{{\bm{B}}}

\def\mD{{\bm{D}}}
\def\mE{{\bm{E}}}
\def\mF{{\bm{F}}}
\def\mG{{\bm{G}}}
\def\mH{{\bm{H}}}

\def\mM{{\bm{M}}}

\def\mP{{\bm{P}}}

\def\mS{{\bm{S}}}

\def\mW{{\bm{W}}}

\usepackage{algpseudocode}
\usepackage{algorithm}


\PassOptionsToPackage{hyphens}{url}
\usepackage[colorlinks = true,
            linkcolor = blue,
            urlcolor  = blue,
            citecolor = blue,
            anchorcolor = blue]{hyperref}
\usepackage{etoolbox}

\usepackage{cite}
\renewenvironment{abstract}
  {{\bfseries\noindent{\abstractname}\par\nobreak}\footnotesize}
  {\bigskip}

\titlespacing{\section}{0pt}{*3}{*1}

\usepackage{authblk}

\usepackage{graphicx}
\usepackage[space]{grffile}
\usepackage{latexsym}
\usepackage{textcomp}
\usepackage{longtable}
\usepackage{tabulary}
\usepackage{booktabs,array,multirow}
\usepackage{amsfonts,amsmath,amssymb}
\providecommand\citet{\cite}
\providecommand\citep{\cite}

\newif\iflatexml\latexmlfalse

\AtBeginDocument{\DeclareGraphicsExtensions{.pdf,.PDF,.png,.png,.png,.png,.tif,.TIF,.jpg,.JPG,.jpeg,.JPEG}}

\usepackage[utf8]{inputenc}
\usepackage[english]{babel}

\usepackage{float}

\iflatexml

\else

\title{\bf A Probabilistic Bayesian Approach to Recover \\$R_2^*$ map and Phase Images for Quantitative Susceptibility Mapping}

\author{\centering Shuai Huang$^1$, James J. Lah$^2$, Jason W. Allen$^1$, and Deqiang Qiu$^1$\thanks{This work is supported by National Institutes of Health under Grants R21AG064405, R01AG072603 and P30AG066511. Corresponding author: Deqiang Qiu (deqiang.qiu@emory.edu).\\ $^1$Department of Radiology and Imaging Sciences, Emory University, Atlanta, GA, 30322, USA\\ $^2$Department of Neurology, Emory University, Atlanta, GA, 30322, USA }}

\date{}

\begin{document}

\vspace{-1em}

\maketitle

Updated final version is accepted and available in ``Magnetic Resonance in Medicine'':\\
 \urlstyle{tt}\url{https://doi.org/10.1002/mrm.29303}
\\

The code files for image reconstruction are available at:\\
\urlstyle{tt}\url{https://github.com/EmoryCN2L/R2Star_Phase_for_QSM}


\newpage

\selectlanguage{english}
\begin{abstract}
\normalsize
{\bfseries Purpose:} Undersampling is used to reduce the scan time for high-resolution 3D magnetic resonance imaging. In order to achieve better image quality and avoid manual parameter tuning, we propose a probabilistic Bayesian approach to recover $R_2^*$ map and phase images for quantitative susceptibility mapping (QSM), while allowing automatic parameter estimation from undersampled data.

\par\null

{\bfseries Theory:} Sparse prior on the wavelet coefficients of images is interpreted from a Bayesian perspective as sparsity-promoting distribution. A novel nonlinear approximate message passing (AMP) framework that incorporates a mono-exponential decay model is proposed. The parameters are treated as unknown variables and jointly estimated with image wavelet coefficients.

\par\null

{\bfseries Methods:} Undersampling takes place in the y-z plane of k-space according to the Poisson-disk pattern. Retrospective undersampling is performed to evaluate the performances of different reconstruction approaches, prospective undersampling is performed to demonstrate the feasibility of undersampling in practice.

\par\null

{\bfseries Results:} The proposed AMP with parameter estimation (AMP-PE) approach successfully recovers $R_2^*$ maps and phase images for QSM across various undersampling rates. It is more computationally efficient, and performs better than the state-of-the-art $l_1$-norm regularization (L1) approach in general, except a few cases where the L1 approach performs as well as AMP-PE.

\par\null

{\bfseries Conclusion:} AMP-PE achieves better performance by drawing information from both the sparse prior and the mono-exponential decay model. It does not require parameter tuning, and works with a clinical, prospective undersampling scheme where parameter tuning is often impossible or difficult due to the lack of ground-truth image.
\end{abstract}%


\sloppy

\pagebreak

\section{Introduction}
\label{sec:intro}
In quantitative magnetic resonance imaging (MRI), we can use multi-echo gradient echo (GRE) sequences to measure tissue properties such as initial magnetization, $T_1$ and $T_2^*$ relaxation rates, and susceptibility differences \cite{Bernstein:2004}. These quantitative values provide direct and precise mappings of tissue properties, and can be used to detect and monitor small pathological changes. In particular, $R_2^*$ map (i.e., the reciprocal of $T_2^*$ map) and quantitative susceptibility mapping (QSM) \cite{Mamisch:T2Star:2012,Wang:QSM:2015,Langkammer:QSM:2013,Deistung:QSM_R2Star:2013,Barbosa:QSM_R2Star:2015,Betts:QSM_R2Star:2016,Qiu1085} are widely used to study iron deposition in the brain \cite{Ordidge:1994,McNeill:2008,Langkammer:QSM_iron:2012,Schweser:QSM:2012,Li:QSM:2011} or pathology such as hemorrhage \cite{Fazekas:1999,Kinoshita:2000,ORegan:2009,Zhang:QSM_hemorrhage:2018,Sun:QSM_hemorrhage:2018} and calcification \cite{Yamada:1996,Gupta:2001,Deistung:QSM_calcification:2013,Chen:QSM_calcification:2014}, etc. In order to accurately characterize the local variabilities of tissue, we need high-resolution 3D volumetric scans that could take $30\sim40$ minutes to acquire fully sampled data in the $k$-space. The long scan time causes discomfort to patients and could introduce motion artifacts to reconstructed images. In this case undersampling is a direct and effective way to reduce the scan time. Parallel imaging methods achieve this goal by exploring the redundancy in measurements from multi-channel receiver coils \cite{Pruessmann:SENSE:1999,Griswold:GRAPPA:2002,Uecker:ESPIRiT:2014}. 

On the other hand, undersampling leads to decreased image quality. We shall rely on prior knowledge about the images to fill in the missing information. For example, images are approximately sparse in some proper basis like the wavelet basis. Most of the wavelet coefficients of an image are close to zero, and the signal energy is concentrated within a small percentage of significant entries. Compressive sensing (CS) methods exploit such sparse prior to improve the image quality \cite{RUP06,CS06,Block:ModelT2:2009,Zhao:MR_mapping:2014,Tamir:T2:2017}. When the sparse prior is enforced through regularization functions such as the $l_1$-norm \cite{l1stable06,Yang2010ARO}, the regularization parameters need to be manually tuned to balance the tradeoff between the data fidelity term and the regularization function. However, parameter tuning is time consuming, and the parameters tuned on a training set might suffer the overfitting problem \cite{Tetko:Overfitting:1995,Hawkins:Overfitting:2004}. The L-curve method has been used to select an empirical regularization parameter \cite{Hansen:l_curve:2000}. Another empirical method was proposed in \cite{Srivastava:2016:wave_denoise} to calculate the denoising thresholds based on statistical summary of 1D electron spin resonance signals, however, it could not be used for denoising 3D MR images that have different statistical properties. Data-driven approaches have also been used to perform CS recovery from undersampled data, and show comparable results to empirically tuned approaches \cite{Khare:MRM:2012,Ahmad:TCI:2015}. Alternatively, the sparse prior can be interpreted from a Bayesian perspective: the signal of interest is assumed to be generated from a sparsity-promoting distribution such as the Laplace distribution. The distribution parameters can be estimated jointly with the sparse signal using approximate message passing (AMP) \cite{Rangan:GAMP:2011,PE_GAMP17}, which makes the AMP approach a better choice in this regard (without the need for exhaustive manual parameter tuning).

AMP is widely used for sparse signal recovery due to its computational efficiency and state-of-the-art performance \cite{Donoho:AMP:2009,Baron:2010,Rangan:GAMP:2011}, it can be formulated either in a denoising form \cite{Guo:SURE:2015,Metzler:Denoising:2016,Ma:AMP_Denoise:2016} or in a Bayesian form \cite{Rangan:GAMP:2011,Krzakala:2012:1}. With its Bayesian formulation we can estimate the parameters by treating them as random variables and maximizing their posteriors \cite{PE_GAMP17}. This is much simpler compared to other approaches that maximize either the likelihood \cite{Vila:EMGM:2013,Kamilov:PE:2014} or the Beth free entropy \cite{Krzakala:2012:1,Krzakala:2012:2}. AMP was originally developed for linear systems \cite{Donoho:AMP:2009,Rangan:GAMP:2011}, the standard AMP has been used with success to recover MR images from linear k-space measurements \cite{Ziniel:DCS:2013,Millard:AMP_MRI:2020,Qiao:AMP_MRI:2020}. Rich et al. \cite{Rich:AMP_PC_MRI:2016,Rich:4Dflow:2018,Aaron:4Dflow:2020} later designed a nonlinear AMP framework for phase-contrast MRI and 4D flow imaging.

Since MR signal intensities at different echo times follow the nonlinear mono-exponential decay model, the standard linear AMP could not be used to recover the $R_2^*$ map. In this paper we propose a new nonlinear AMP framework that incorporates the mono-exponential decay model, and use it to recover the $R_2^*$ map and complex multi-echo images. QSM can then be computed from the complex multi-echo images subsequently \cite{Liu:MEDI:2013,Liu:MEDI:2012,Qiu1085}. Compared to regularization approaches that require parameter tuning, our proposed approach automatically and adaptively estimates the parameters with respect to each dataset. By drawing additional information from the mono-exponential decay model, it achieves better performance and offers a convenient way to recover $R_2^*$ maps and phase images for QSM from undersampled data.

\section{Theory}
\subsection{Problem Formulation}
\label{sec:problem_formulation}
\begin{figure}[tbp]
\centering
\subfigure[]{
\label{fig:k_space_sampling}
\includegraphics[height=.22\textwidth]{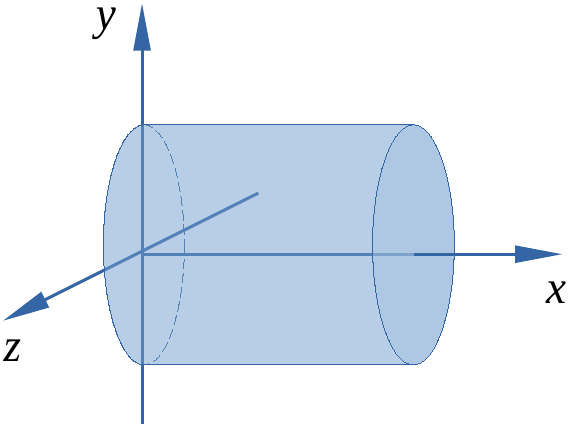}
\includegraphics[height=.22\textwidth]{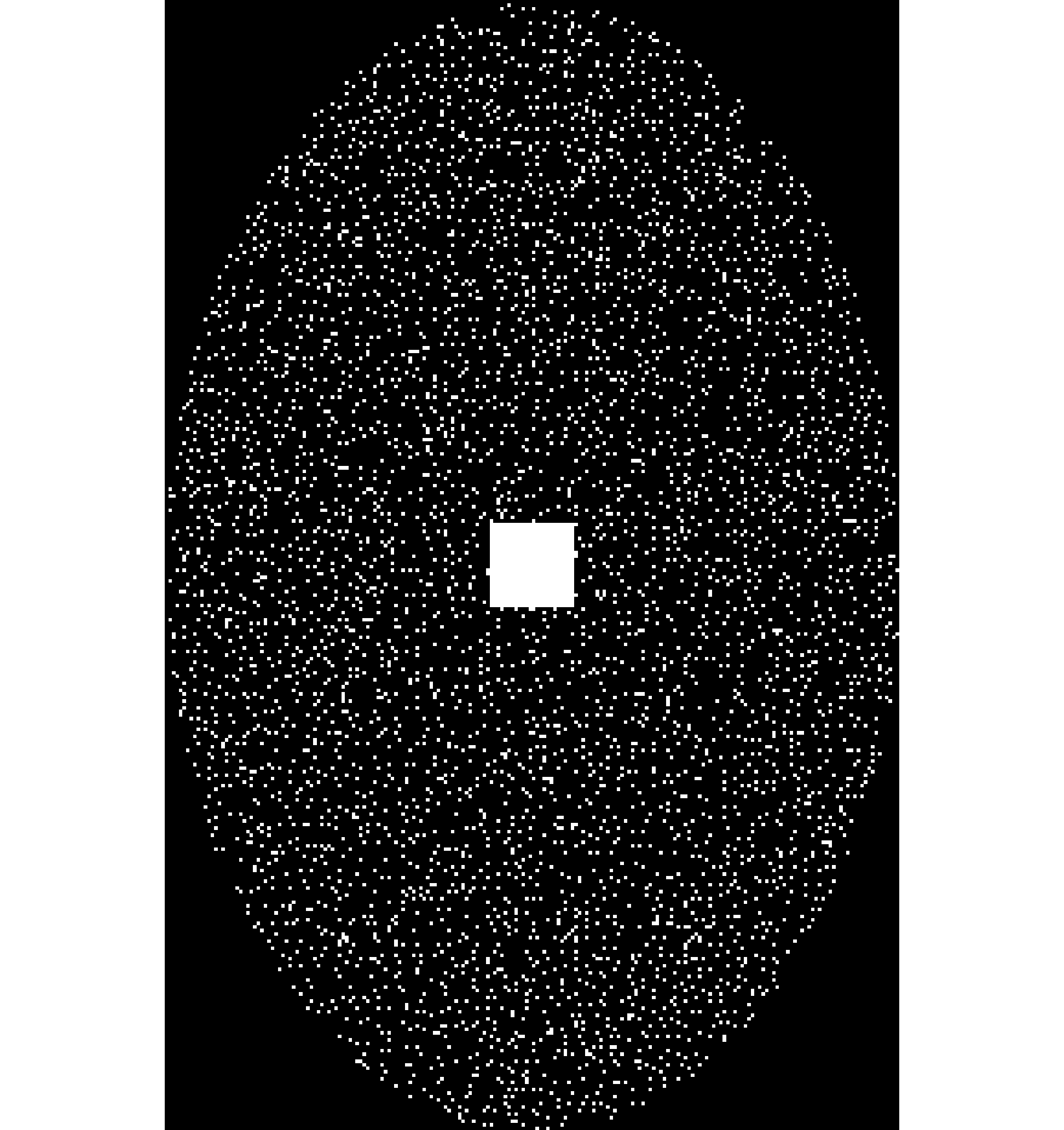}}
\subfigure[]{
\label{fig:monoexponential_te}
\includegraphics[height=.22\textwidth]{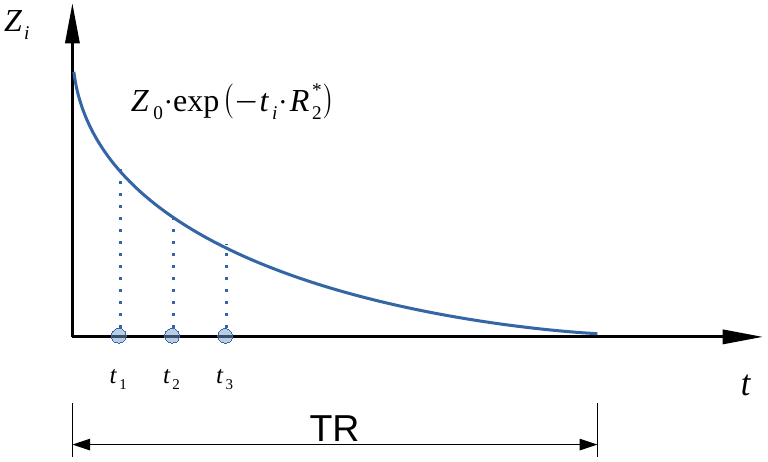}}\\
\subfigure[$\vz_0$]{
\label{fig:2d_brain_ori}
\includegraphics[width=.235\textwidth]{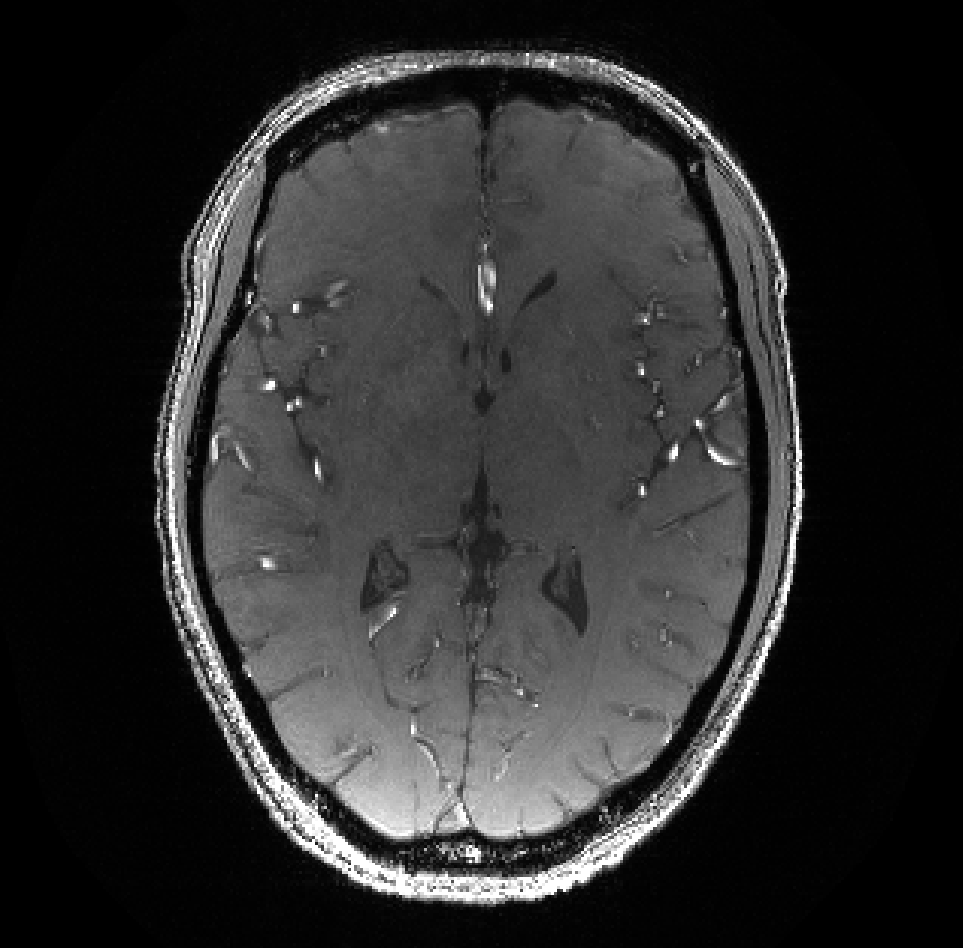}}
\subfigure[$|\vv_0|$]{
\label{fig:2d_brain_wavelet}
\includegraphics[width=.235\textwidth]{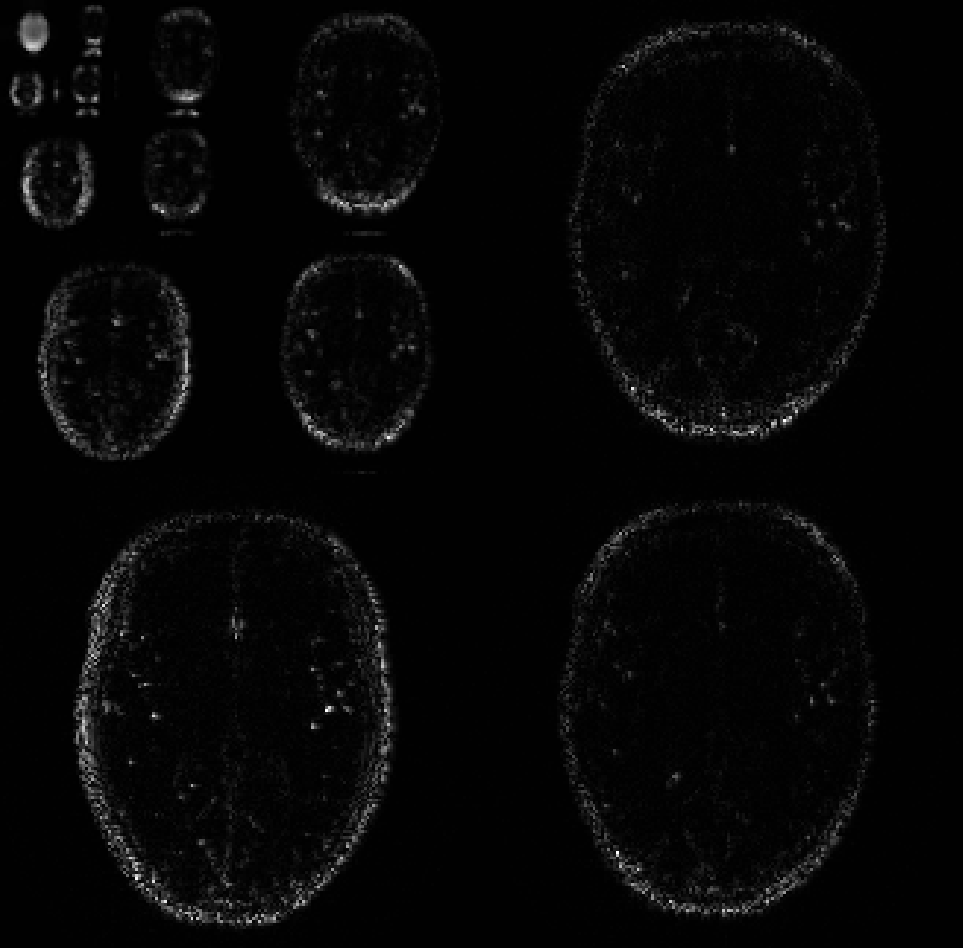}}
\subfigure[$\widehat{\vz}_0$]{
\label{fig:2d_brain_rec}
\includegraphics[width=.235\textwidth]{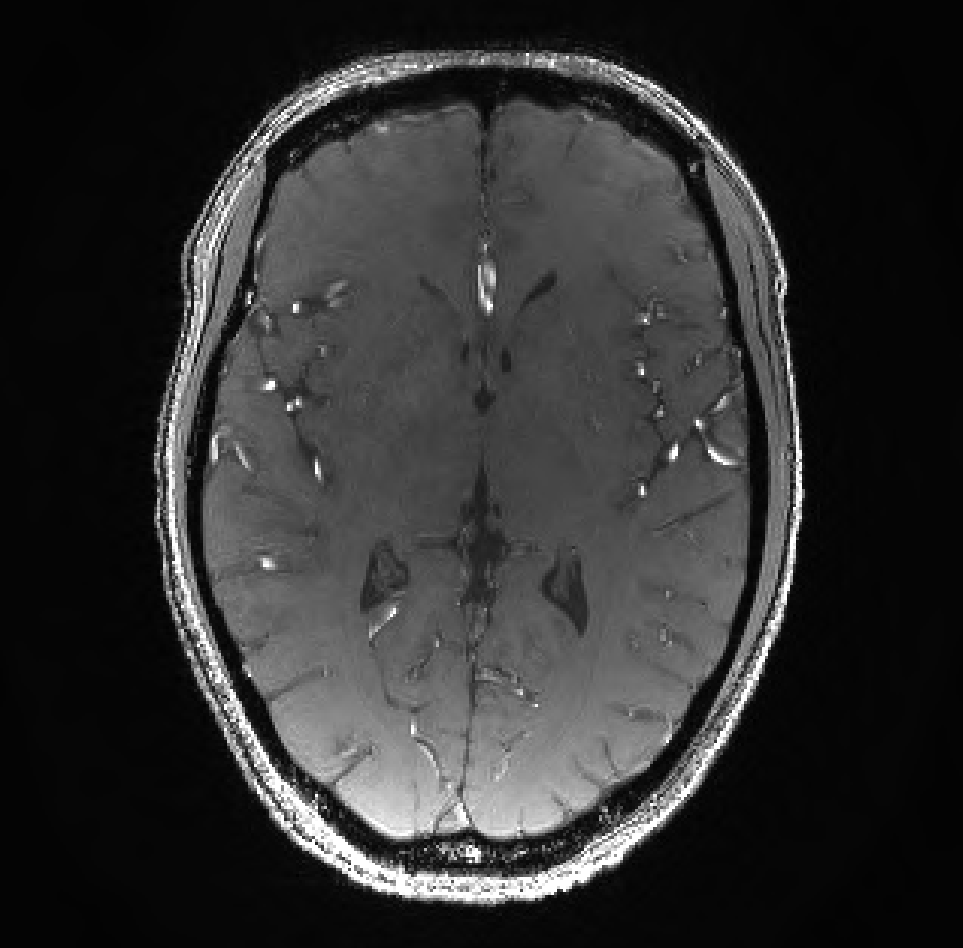}}
\subfigure[$|\vv_i|$]{
\label{fig:magnitude_complex_wavelet}
\includegraphics[width=.235\textwidth]{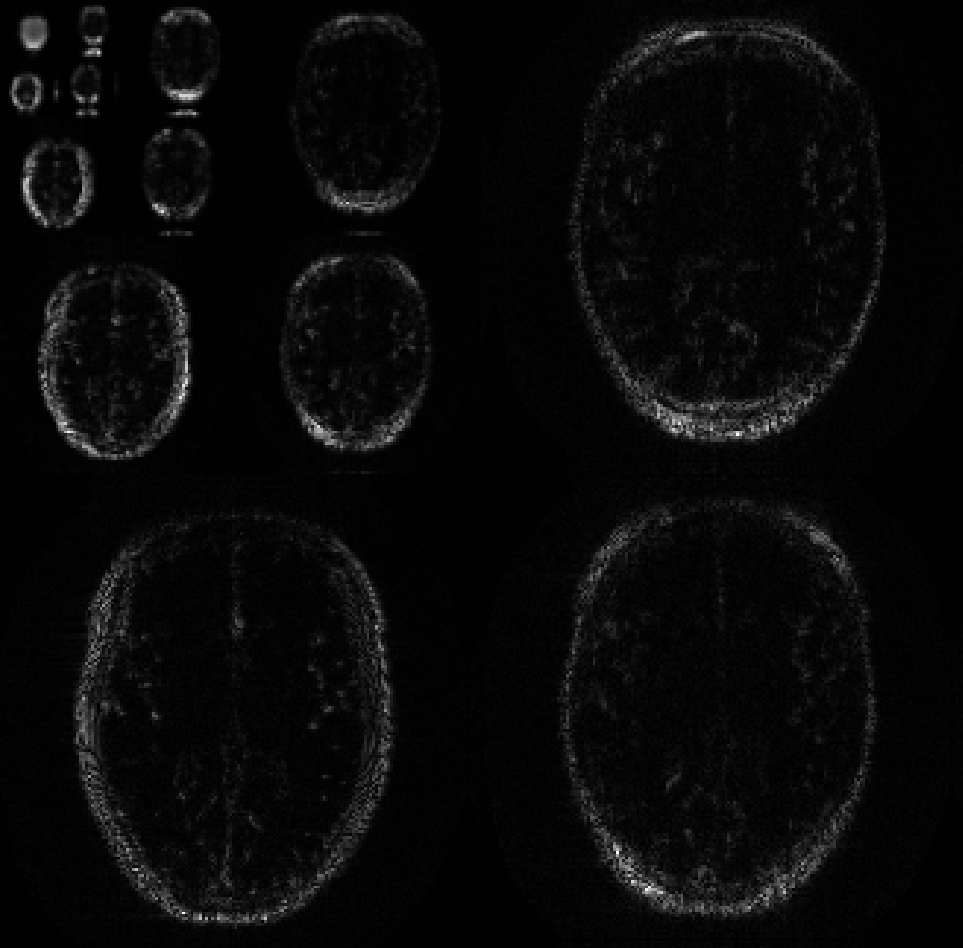}}
\caption{Undersampling acquisition of a GRE sequence: (a) The 3D $k$-space is undersampled to reduce the scan time; (b) Multi-echo $k$-space data acquired at different echo times are needed for the recovery of $R_2^*$ map and QSM. Sparse prior on the wavelet coefficients: (c) The initial magnetization image $\vz_0$; (d) The magnitude of its sparse wavelet coefficients $|\vv_0|$; (e) Reconstructed image $\widehat{\vz}_0$ using the top $20\%$ wavelet coefficients (NRMSE=0.024); (f) The magnitude of complex wavelet coefficients $|\vv_i|$ of a complex multi-echo image $\vz_i$.} 
\end{figure}

As shown in Fig. \ref{fig:k_space_sampling}, undersampling in the $k$-space of a 3D acquisition takes place along the two phase-encoding directions $y$ and $z$, whereas the readout direction $x$ is fully sampled. The elliptical Poisson-disk sampling pattern is adopted to select the sampling locations in the $y$-$z$ plane. It imposes a minimum pairwise-distance constraint between any two sampling locations, thus producing a more uniform sampling distribution than the usual random sampling. Furthermore, as shown in Fig. \ref{fig:monoexponential_te}, the $k$-space data are independently acquired at multiple echo times (TE) within one repetition time (TR) of a gradient-echo sequence (GRE). The magnetization across different TEs at every voxel can be modeled by the mono-exponential decay \cite{MRI:Nishimura:2010}, i.e. the magnitude of the \emph{complex} multi-echo image $\vz_i$ at echo time $t_i$ is
\begin{align}
\label{eq:mono_exp_decay_model}
    |\vz_i|=\vz_0\cdot\exp\left(-t_i\cdot \vr_2^*\right),\quad i\in\{1,\cdots,I\}\,,
\end{align}
where $\vz_0$ is the initial magnetization image, and $\vr_2^*$ is the effective transverse $R_2^*$ relaxation-rate map. Multiple receiver coils can be used to acquire measurements to improve the overall SNR and image quality. Let $\vy_{i}$ denote all the multi-coil measurements at time $t_i$, and $\vw_{i}$ denote the measurement noise. We then have
\begin{align}
\label{eq:multi_echo_measurement_model}
\begin{split}
    \vy_{i}&=\mP_i\mF\mS\vz_i+\vw_{i}\\
    &=\mA_i\vz_i+\vw_{i}\,,
\end{split}
\end{align}
where $\mP_i$ is the undersampling matrix at time $t_i$, $\mF$ is the Fourier operator, the diagonal matrix $\mS$ contains the sensitivity maps of all receiver coils, and $\mA_{i}=\mP_i\mF\mS$ is the resulting measurement matrix at time $t_i$. When the central $k$-space is fully sampled as shown in Fig. \ref{fig:k_space_sampling}, the sensitivity maps can be estimated using the ESPIRiT approach \cite{Uecker:ESPIRiT:2014}. We shall first recover the $\vr_2^*$ map, the initial magnetization $\vz_0$ and the complex-valued multi-echo images $\vz_i$ from the measurements $\vy_{i}$, and then compute QSM from $\vz_i$.

As shown in Fig. \ref{fig:2d_brain_ori}-\ref{fig:2d_brain_rec}, the initial magnetization image $\vz_0$ is approximately sparse in the wavelet basis \cite{DBWav92}: most of wavelet coefficients are close to zero, and the image can be reconstructed well using only the significant coefficients. When it comes to the complex multi-echo image $\vz_i$, the complex wavelet coefficients $\vv_i$ are also sparse (see Fig. \ref{fig:magnitude_complex_wavelet}). This allows us to make use of the sparse prior on images to improve the recovery performance. Let $\mH$ denote the wavelet transform operator, the recovery problem in this paper is then formulated with respect to the wavelet coefficients $\vv_i,\vv_0$ of $\vz_i,\vz_0$ instead:
\begin{align}
    \label{eq:wavelet_xi}
    \vv_i &= \mH\vz_i\\
    \label{eq:wavelet_x0}
    \vv_0 &= \mH\vz_0\,.
\end{align}

From a Bayesian perspective, we assume the wavelet coefficients $\vv$ follow the Laplace distribution that produces sparse signals, and that they are identically and independently distributed (i.i.d.):
\begin{align}
\label{eq:laplace_dist}
    p(v|\lambda)= \frac{1}{2}\lambda\cdot\exp(-\lambda|v|)\,,
\end{align}
where $\lambda>0$ is the distribution parameter. The measurement noise $\vw$ can be modeled as i.i.d. additive white Gaussian noise (AWGN):
\begin{align}
\label{eq:awgn}
    p(w|\theta) =\mathcal{N}(w|0,\theta^2)\,,
\end{align}
where the mean is $0$, and $\theta$ is the standard deviation of noise. Given the noisy measurements $\vy$, the recovered wavelet coefficients $\widehat{\vv}$ can be obtained through max-sum approximate message passing (AMP) \cite{Rangan:GAMP:2011}:
\begin{align}
    \label{eq:map_v_est}
    \widehat{v}=\arg\max_v\ p(v|\vy)\,.
\end{align}
By treating the distribution parameters ${\lambda,\theta}$ as random variables, we can compute their maximum-a-posteriori (MAP) estimations as well \cite{PE_GAMP17}.
\begin{align}
    \widehat{\lambda}&=\arg\max_{\lambda}\ p(\lambda|\vy,\widehat{v})\\
    \widehat{\theta}&=\arg\max_{\theta}\ p(\theta|\vy,\widehat{v})\,.
\end{align}

When the measurement matrix contains i.i.d zero-mean Gaussian entries, the convergence behavior of AMP in the large system limit can be guaranteed and characterized by state evolution analysis \cite{Donoho:AMP:2009,Bayati:SE:2011}. In the case of the MRI measurement model in \eqref{eq:multi_echo_measurement_model}, the measurement matrix $\mA_i\mH^{-1}$ with respect to $\vv_i$ is not a random Gaussian matrix: it consists of the undersampling operator $\mP_i$, the Fourier operator $\mF$, the sensitivity maps $\mS$ and the inverse wavelet operator $\mH^{-1}$. Although establishing state evolution analysis for generic measurement matrices is still an open question, the damping and mean removal operations are able to stabilize and ensure the convergence of AMP \cite{Rangan:DampingCvg:2014,Vila:DampingMR:2015}.

In order to further improve the recovery performance, we need to combine information from the multi-echo measurement model in \eqref{eq:multi_echo_measurement_model} and the mono-exponential decay model in \eqref{eq:mono_exp_decay_model}. In the following we first compute the distribution $p_{\mathcal{M}}(\vz_i|\vy)$ of multi-echo images $\vz_i$ based on the multi-echo measurement model alone. We then integrate $p_{\mathcal{M}}(\vz_i|\vy)$ into the mono-exponential decay model to recover the multi-echo images $\vz_i$, the initial magnetization $\vz_0$ and the $R_2^*$ map $\vr_2^*$. Compared to $p_{\mathcal{M}}(\vz_i|\vy)$, the combined posterior distribution $p(\vz_i|\vy)$ we use to recover $\vz_i$ also contains information from the mono-exponential decay model, which thus leads to better performances. In particular, since the sparse priors are imposed on the wavelet coefficients $\vv_i,\vv_0$ of the images $\vz_i,\vz_0$, the recovery problem is then formulated with respect to $\vv_i,\vv_0$ when we need to invoke the sparsity-promoting distributions $p(\vv_i|\lambda_i)$ and $p(\vv_0|\lambda_0)$.

\subsection{Multi-echo Image Distribution}
\begin{figure}[tbp]
\begin{center}
\subfigure[]{
\label{fig:factor_graph_multi_echo}
\includegraphics[height=.4\textwidth]{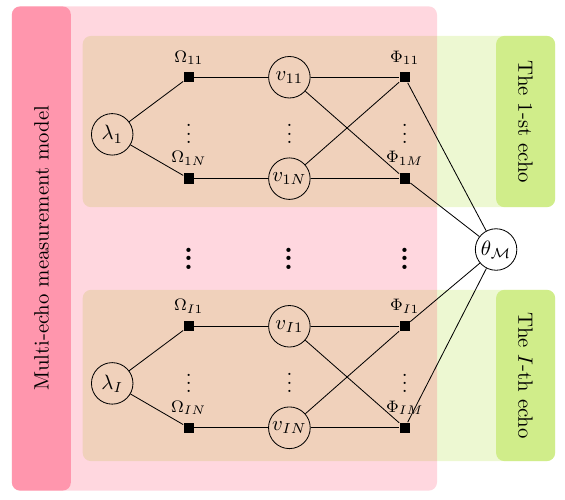}}\\
\subfigure[]{
\label{fig:factor_graph_r2star_multi_echo}
\includegraphics[width=.83\textwidth]{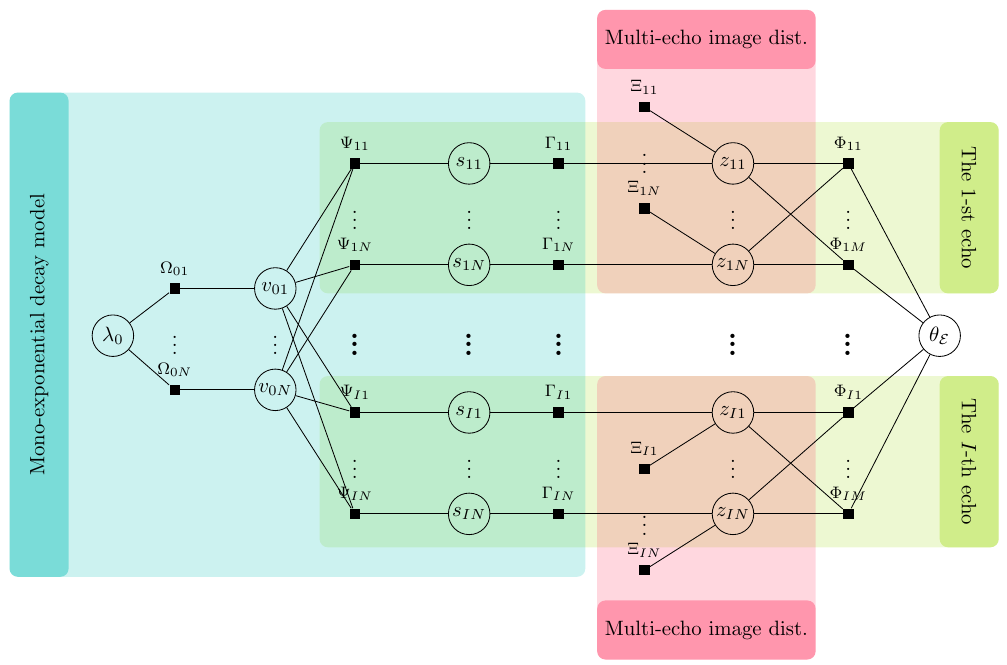}}
\end{center}
\caption{(a) The factor graph used to compute the multi-echo image distribution $p_{\mathcal{M}}(\vz_i|\vy)$ from the multi-echo measurement model in \eqref{eq:multi_echo_measurement_model}. (b) The factor graph used to recover $\vr_2^*$, $\vz_0$ and $\vz_i$ by combining the multi-echo measurement model in \eqref{eq:multi_echo_measurement_model} with the mono-exponential decay model in \eqref{eq:mono_exp_decay_model}.}
\label{fig:factor_graph_amp_pe}
\end{figure}

The Bayesian model used to calculate the multi-echo image distribution $p_{\mathcal{M}}(\vv_i|\vy)$ and, by extension, $p_{\mathcal{M}}(\vz_i|\vy)$ is given by the factor graph shown in Fig. \ref{fig:factor_graph_multi_echo}. The variable nodes are represented by ``$\bigcirc$'' and contain random variables in the Bayesian model, the factor nodes are represented by ``$\blacksquare$'' and encode probability distributions of the variables. Messages about how the variables are distributed are passed among the nodes in the factor graph. During the message passing process, the variable node simply passes the messages it receives to the factor nodes. Whereas the factor node first combines the message it encodes with the messages it receives, and then passes the combined message to the variable nodes.

We use the following notations for the messages between the $n$-th variable node $v_{in}$ and the $m$-th factor node $\Phi_{im}$ in the $i$-th echo:
\begin{itemize}
    \item $\Delta_{v_{in}\rightarrow\Phi_{im}}$ denotes the message from $v_{in}$ to $\Phi_{im}$,
    \item $\Delta_{\Phi_{im}\rightarrow v_{in}}$ denotes the message from $\Phi_{im}$ to $v_{in}$,
\end{itemize}
where $i\in\{1,\cdots,I\}$, $n\in\{1,\cdots,N\}$, and $m\in\{1,\cdots,M\}$. Both $\Delta_{v_{in}\rightarrow\Phi_{im}}$ and $\Delta_{\Phi_{im}\rightarrow v_{in}}$ are functions of $v_{in}$, and they are expressed in the ``$\log$'' domain in this paper. The messages will be passed among the nodes iteratively until a consensus on how the variables are distributed is reached \cite{Kschischang:2001,Koller:2009}. Detailed expressions of the messages are given in Appendix \ref{subsec:messages_multi_echo_image_distribution}.

The signal prior distribution parameter $\lambda_i$ can be estimated by maximizing its posterior \cite{PE_GAMP17}:
\begin{align}
    \hat{\lambda}_i=\arg\max_{\lambda_i}\ p(\lambda_i|\vy)=\arg\max_{\lambda_i}\ \sum_{n}\Delta_{\Omega_{in}\rightarrow\lambda_i}\,.
\end{align}

The noise distribution parameter $\theta_{\mathcal{M}}$ can also be estimated by maximizing its posterior:
\begin{align}
    \hat{\theta}_{\mathcal{M}}=\arg\max_{\theta_{\mathcal{M}}}\ p(\theta_{\mathcal{M}}|\vy)=\arg\max_{\theta_{\mathcal{M}}}\ \sum_{im}\Delta_{\Phi_{im}\rightarrow\theta_{\mathcal{M}}}\,.
\end{align}

The distributions in AMP are approximated by Gaussian distributions to simplify the message passing process \cite{Minka:2001,Minka:Div:2005}. The distribution $p_{\mathcal{M}}(\vv_i|\vy)$ from the multi-echo measurement model is then
\begin{align}
\label{eq:multi_echo_wave}
\begin{split}
    p_{\mathcal{M}}(v_{in}|\vy)\ &\propto\exp\Big(\Delta_{\Omega_{in}\rightarrow v_{in}}+\sum_k\Delta_{\Phi_{ik}\rightarrow v_{in}}\Big)\\
    &\approx\mathcal{N}\big(v_{in}\ \left|\ {\mu}_{in}(v),\ {\tau}_{in}(v)\right.\big)\,,
\end{split}
\end{align}
where ${\mu}_{in}(v)$ and ${\tau}_{in}(v)$ are the mean and variance of the Gaussian approximation. Let $\vh^{-1}_n$ denote the $n$-th column of the inverse wavelet operator $\mH^{-1}$, and $\|\mH^{-1}\|_F$ denote its Frobenius norm. Under the i.i.d. assumption of wavelet coefficients in $\vv_i$, the distribution $p_{\mathcal{M}}(\vz_i|\vy)$ can be calculated straightforwardly:
\begin{align}
\label{eq:multi_echo_image_prior}
    p_{\mathcal{M}}(z_{in}|\vy)=\mathcal{N}\big(z_{in}\ \left|\ {{\mu}_{\mathcal{M}}}_{in}(z),\ {{\tau}_{\mathcal{M}}}_{in}(z)\right.\big)\,,
\end{align}
where ${{\mu}_{\mathcal{M}}}_{in}(z)=\left\langle\vh^{-1}_n,\boldsymbol{\mu}_{i}(v)\right\rangle$ and ${{\tau}_{\mathcal{M}}}_{in}(z)=\frac{1}{N}\|\mH^{-1}\|_F^2\cdot{\tau}_{in}(v)$.

\subsection{Proposed Nonlinear AMP Framework}
By treating the $R_2^*$ map $\vr_2^*$ as the ``model'' parameters to be estimated, we can rewrite the original mono-exponential decay model in \eqref{eq:mono_exp_decay_model} as follows
\begin{align}
    |\vz_i|=\mB_i(\vr_2^*)\cdot\vz_0=\mB_i(\vr_2^*)\cdot\mH^{-1}\vv_0\,,
\end{align}
where $\mB_i(\vr_2^*)$ is a diagonal matrix whose diagonal entries are $\exp(-t_i\cdot\vr_2^*)$. The distribution $p_{\mathcal{M}}(\vz_i|\vy)$ from the multi-echo measurement model can be integrated into the mono-exponential decay model via the factor node $\Xi_{in}$ of the factor graph in Fig. \ref{fig:factor_graph_r2star_multi_echo}. We have that
\begin{align}
    \Xi(z_{in})=p_{\mathcal{M}}(z_{in}|\vy)\,.
\end{align}
Detailed expressions of the messages exchanged between the variable and factor nodes are given in Appendix \ref{subsec:messages_t2star_map}.

The signal prior distribution parameter $\lambda_0$ can be estimated by maximizing its posterior:
\begin{align}
    \hat{\lambda}_0=\arg\max_{\lambda_0}\ p(\lambda_0|\vy) = \arg\max_{\lambda_0}\ \sum_d\Delta_{\Omega_{0d}\rightarrow\lambda_0}\,.
\end{align}

The noise distribution parameter $\theta_{\mathcal{E}}$ can also be estimated by maximizing its posterior:
\begin{align}
    \hat{\theta}_{\mathcal{E}}=\arg\max_{\theta_{\mathcal{E}}}\ p(\theta_{\mathcal{E}}|\vy)=\arg\max_{\theta_{\mathcal{E}}}\ \sum_{ik}\Delta_{\Phi_{ik}\rightarrow\theta_{\mathcal{E}}}\,.
\end{align}

The $R_2^*$ map $\vr_2^*$ and the initial magnetization $\vz_0$ can be recovered using \eqref{eq:recover_r2star},\eqref{eq:recover_proton} in Appendix \ref{subsec:messages_t2star_map}. For the recovery of multi-echo image $\vz_i$, we need to combine the message $\Xi(z_{in})=p_{\mathcal{M}}(z_{in}|\vy)$ in \eqref{eq:multi_echo_image_prior} from the multi-echo measurement model with messages from the mono-exponential decay model. We then have:
\begin{align}
    \hat{z}_{in} = \arg\max_{z_{in}}\ p(z_{in}|\vy)=\arg\max_{z_{in}}\ \log\Xi(z_{in})+\Delta_{\Gamma_{in}\rightarrow z_{in}}+\sum_{ik}\Delta_{\Phi_{ik}\rightarrow z_{in}}\,.
\end{align}

We derive the messages in AMP under the GAMP formulation \cite{Rangan:GAMP:2011}, and compute the MAP estimations of distribution parameters according to \cite{PE_GAMP17}. To simplify the notations, we use $\mE$ and $\mG_i$ to denote the following measurement operators
\begin{gather}
    \mE=\left[\begin{array}{c}\mB_1(\vr_2^*)\mH^{-1}\\ \vdots\\\mB_I(\vr_2^*)\mH^{-1}  \end{array}\right]\\
    \mG_i=\mA_i\mH^{-1}\,.
\end{gather}
The mono-exponential decay model in \eqref{eq:mono_exp_decay_model} and the multi-echo measurement model in \eqref{eq:multi_echo_measurement_model} can then be rewritten with respect to the wavelet coefficients $\vv_0,\vv_i$:
\begin{gather}
    \left[\begin{array}{c} |\vz_1|\\ \vdots\\ |\vz_I| \end{array}\right]=\mE\vv_0\\
    \vy_i=\mG_i\vv_i+\vw_i\,.
\end{gather}

Let $\|\mE\|_F$ and $\|\mG_i\|_F$ denote the Frobenius norms of $\mE,\mG_i$ respectively. The AMP algorithm to calculate the multi-echo image distribution $p_{\mathcal{M}}(z_{in}|\vy)$ is summarized in Algorithm S1 of the Supporting Information, and the AMP algorithm to recover $\vr_2^*,\vz_0,\vz_i$ is summarized in Algorithm S2 of the Supporting Information. QSM can then be computed from the complex multi-echo images $\{\vz_i|_{i=1}^I\}$.

\section{Methods}
We acquired \emph{in vivo} 3D brain data on a 3T MRI scanner (Prisma model, Siemens Healthcare, Erlangen, Germany), with written consent obtained from the subjects before imaging under the approval from the Institutional Review Board of Emory University. The data were acquired with a 32-channel head coil using the GRE sequence. The sensitivity maps of the 32 coils are estimated from the data by extending the 2D ESPIRiT approach in \cite{Uecker:ESPIRiT:2014} to the 3D case. In order to reduce the scan time down to around 10 minutes, we are interested in the low-sampling-rate regime where the undersampling rates vary in $\{10\%,\ 15\%,\ 20\%\}$. Retrospective and prospective undersampling schemes were adopted in the experiments. The retrospective scheme acquires a fully-sampled dataset during the scan, and then undersamples the dataset retrospectively. It provides the ground-truth image and is used to compare different approaches. Since the prospective scheme acquires the undersampled dataset in real time, it is used to validate the feasibility of performing undersampling in practice. The code files for reconstructing the images are available at \urlstyle{tt}\url{https://github.com/EmoryCN2L/R2Star_Phase_for_QSM}

\paragraph{Retrospective Undersampling:} In order to provide the ground-truth reference for evaluating different approaches, the k-space was fully sampled within an elliptical region in the $y-z$ plane as shown in Fig. \ref{fig:k_space_sampling}. The retrospective undersampling took place in the $y-z$ plane afterwards according to randomly generated Poisson-disk sampling patterns as shown in Fig. \ref{fig:k_space_sampling}, whereas the readout $x$-direction was always fully sampled at each TE. The minimum distance between any two sampling locations is set to 2 pixels for best performance. Two acquisition protocols were used here, seven subjects were recruited for the first protocol, and five subjects were recruited for the second protocol. For each protocol, one of the subject was used as the training dataset and the other subjects were used as the test datasets.
\begin{itemize}
\item \emph{Protocol 1 (P1-R):} We have the flip angle = $15$\textdegree, the number of echoes = 4, the first echo time = 7.32 ms, echo spacing = 8.68 ms, slice thickness = 0.6875 mm, in-plane resolution = 0.6875 mm $\times$ 0.6875 mm, bandwidth per pixel = 260 Hz, TR = 38 ms, and FOV = 220 mm $\times$ 220 mm. The acquisition time is 33 minutes.
\item \emph{Protocol 2 (P2-R):} We have the flip angle = $15$\textdegree, the number of echoes = 4, the first echo time = 7.91 ms, echo spacing = 9.19 ms, slice thickness = 0.6875 mm, in-plane resolution = 0.6875 mm $\times$ 0.6875 mm, bandwidth per pixel = 260 Hz, TR = 41 ms, and FOV = 220 mm $\times$ 220 mm. The acquisition time is 35 minutes.
\end{itemize}

\paragraph{Prospective Undersampling:} The prospective protocols were implemented via pulse sequence programming using the ``IDEA'' platform from Siemens. The undersampling took place in the $y-z$ plane in real time, and the readout $x$-direction was always fully sampled. Two acquisition protocols were used to validate the prospective scheme. Three subjects were recruited for the first protocol, and four subjects were recruited for the second protocol.
\begin{itemize}
    \item \emph{Protocol 1 (P1-P):} We have the flip angle = $15$\textdegree, the number of echoes = 4, the first echo time = 7.32 ms, echo spacing = 8.68 ms, slice thickness = 0.6875 mm, in-plane resolution = 0.6875 mm $\times$ 0.6875 mm, bandwidth per pixel = 260 Hz, TR = 38 ms, and FOV = 220 mm $\times$ 220 mm. When the undersampling rates vary in $\{10\%,\ 15\%,\ 20\%,\ 100\%\}$, the acquisition times are 4.23, 6.32, 8.43 and 33 minutes respectively. 
    \item \emph{Protocol 2 (P2-P):} We have the flip angle = $15$\textdegree, the number of echoes = 4, the first echo time = 7.91 ms, echo spacing = 9.19 ms, slice thickness = 0.6875 mm, in-plane resolution = 0.6875 mm $\times$ 0.6875 mm, bandwidth per pixel = 260 Hz, TR = 41 ms, and FOV = 220 mm $\times$ 220 mm. When the undersampling rates vary in $\{10\%,\ 15\%,\ 20\%,\ 100\%\}$, the acquisition times are 6.55, 9.8, 13.07 and 35 minutes respectively. 
\end{itemize}

The Daubechies wavelet family is chosen to obtain the sparse representation of an image \cite{DBWav92}. The orthogonal ``db1-db10'' wavelet bases are commonly used, and the complexity of the wavelet basis increases with respect to its order. For the reconstructions of $R_2^*$ map and QSM, we observe that using a higher order wavelet basis generally produces better image quality. In the experiments, we use the db6 basis with 4 levels to balance the tradeoff between wavelet complexity and image quality. 

\subsection{Reconstruction Approaches}
We compare the proposed ``AMP with parameter estimation'' (AMP-PE) approach with the baseline least squares (LSQ) approach and the state-of-the-art $l_1$-norm regularization (L1) approach \cite{Yang2010ARO}. 

\begin{itemize}
    \item The least squares approach:
    \begin{subequations}
    \begin{align}
        &\min_{\vz_1,\cdots,\vz_I}\ \sum_i\|\vy_i-\mA_i\vz_i\|_2^2\\
        &\min_{\vz_0,\vr_2^*}\ \sum_i\Big\||\vz_i|-\vz_0\cdot\exp(-t_i\cdot\vr_2^*)\Big\|_2^2\,.
    \end{align}
    \end{subequations}
    The least squares approach does not require parameter tuning, and the solutions can be obtained using gradient descent. In particular, the recovery of $\vz_0$ and $\vr_2^*$ is performed in an alternating fashion until convergence. When $\vz_0$ is being recovered, $\vr_2^*$ is fixed; conversely, when $\vr_2^*$ is being recovered, $\vz_0$ is fixed. 
    \item The $l_1$-norm regularization approach:
    \begin{subequations}
    \begin{align}
        \label{eq:l1_norm_multi_echo}
        &\min_{\vv_1,\cdots,\vv_I}\ \sum_i\|\vy_i-\mG_i\vv_i\|_2^2+\kappa\cdot\|\vv_i\|_1\\
        &\min_{\vv_0,\vr_2^*}\ \sum_i\Big\||\vz_i|-\mH^{-1}\vv_0\cdot\exp(-t_i\cdot\vr_2^*)\Big\|_2^2+\xi\cdot\|\vv_0\|_1\,,
    \end{align}
    \end{subequations}
    where $\kappa$ and $\xi$ are the regularization parameters. We can choose the parameters in two ways. In the first way referred to as the ``L1-T'' approach, we follow the established practice to tune the parameters on a training set that is acquired under the same condition as the test set \cite{WITTEN2011147}, and then use the tuned parameters on the test set. In the second way referred to as the ``L1-L'' approach, we compute empirical parameters for each test set using the L-curve method. The obtained parameters for retrospective undersampling are given in Table \ref{tab:parameter_1st_protocol}, where $\kappa$ is tuned in an approximate-logarithmic scale from $\{5e^{-4}, 1e^{-3}, 5e^{-3}, 1e^{-2},\cdots, 1, 5, 10, 50\}$ and $\xi$ is tuned in a similar fashion from $\{1e^{-7},5e^{-7},1e^{-6},5e^{-6},\cdots,5e^{-4},1e^{-3},5e^{-3},1e^{-2}\}$. Note that since the ground-truth reference is not available in prospective undersampling schemes, parameter tuning can not be performed. Only the L-curve method is used to compute the parameters in the prospective case, and the parameters are basically the same as those in the retrospective case. After the parameters are selected, the solutions can be solved using FISTA \cite{Beck:FISTA:2009}. The recovery of $\vv_0$ and $\vr_2^*$ is also performed in an alternating fashion until convergence.
    \item For the proposed AMP-PE approach, when the sampling rate is very low ($\sim 10\%$), we need to use the damping operation \cite{Rangan:DampingCvg:2014} to stabilize the AMP update of the wavelet coefficients $\vv$. Let $\mu_d^{(t)}(v)$ denote the damped solution in the previous $t$-th iteration, and $\mu^{(t+1)}(v)$ denote the undamped solution in the $(t+1)$-th iteration. The damping operation simply proceeds as follows:
\begin{align}
    \mu_d^{(t+1)}(v) = \mu_d^{(t)}(v)+\alpha\cdot\left(\mu^{(t+1)}(v)-\mu_d^{(t)}(v)\right)\,,
\end{align}
where $\alpha\in(0,1]$ is the damping rate, $\mu_d^{(t+1)}(v)$ is the damped solution in the $(t+1)$-th iteration. The damping rate $\alpha$ can be considered as step size of this iterative update. When $\alpha$ goes to $0$, the iterative update would stop. When $\alpha=1$, the iterative update directly passes down the undamped solution $\mu^{(t+1)}(v)$ to the next iteration, and no damping operation is performed. When the sampling rate is $10\%$, we choose $\alpha=0.5$ to slow down the iterative update. When the sampling rate is relatively higher ($\geq 15\%$), we can skip the damping step and choose $\alpha=1$.
\end{itemize}

\begin{table}[tbp]
\caption{Retrospective undersampling: parameters in the $l_1$-norm regularization approach. The 1st (S1) and 8th (S8) subjects are used as training data, the rest are used as test data.}
\vspace{0.5em}
\label{tab:parameter_1st_protocol}
\centering
\begin{tabular}{lllccccccc}
\toprule
& & &{\bfseries Parameter tuning } & \multicolumn{6}{c}{\bfseries L-curve} \\ \cline{5-10}  
& & & S1 & S2 & S3 & S4 & S5 & S6 & S7 \\ \midrule
& &$\kappa$ & 0.1 & 0.1 & 0.1 & 0.1 & 0.1 & 0.1 & 0.1  \\ 
& \multirow{-2}{*}{$10\%$} &$\xi$ & $5e^{-5}$ & $5e^{-5}$ & $5e^{-5}$ & $1e^{-5}$ & $5e^{-5}$ & $5e^{-5}$ & $5e^{-5}$ \\ \cline{2-10}
& &$\kappa$ & 0.1 & 0.1 & 0.1 & 0.1 & 0.1 & 0.1 & 0.1  \\ 
& \multirow{-2}{*}{$15\%$} &$\xi$ & $5e^{-5}$ & $1e^{-5}$ & $1e^{-5}$ & $1e^{-5}$ & $5e^{-5}$ & $5e^{-5}$ & $5e^{-5}$ \\ \cline{2-10}
& &$\kappa$ & 0.1 & 0.1 & 0.1 & 0.1 & 0.1 & 0.1 & 0.1  \\ 
\multirow{-6}{*}{P1-R} & \multirow{-2}{*}{$20\%$} &$\xi$ & $5e^{-5}$ & $1e^{-5}$ & $1e^{-5}$ & $1e^{-5}$ & $5e^{-5}$ & $5e^{-5}$ & $5e^{-5}$ \\ \bottomrule

& & &{\bfseries Parameter tuning } & \multicolumn{6}{c}{\bfseries L-curve} \\ \cline{5-10}
& & & S8 & \multicolumn{1}{c}{S9} &\multicolumn{1}{c}{S10} & \multicolumn{1}{c}{S11} & \multicolumn{1}{c}{S12} \\ \midrule
& &$\kappa$ & 0.1 & \multicolumn{1}{c}{0.1} & \multicolumn{1}{c}{0.1} & \multicolumn{1}{c}{0.1} & 0.1  \\ 
& \multirow{-2}{*}{$10\%$} &$\xi$ & $5e^{-5}$ & \multicolumn{1}{c}{$5e^{-5}$} & \multicolumn{1}{c}{$5e^{-5}$} & \multicolumn{1}{c}{$5e^{-5}$} & $5e^{-5}$\\ \cline{2-10}
& &$\kappa$ & 0.1 & \multicolumn{1}{c}{0.1} & \multicolumn{1}{c}{0.1} & \multicolumn{1}{c}{0.1} & 0.1  \\ 
& \multirow{-2}{*}{$15\%$} &$\xi$ & $5e^{-5}$ & \multicolumn{1}{c}{$5e^{-5}$} & \multicolumn{1}{c}{$5e^{-5}$} & \multicolumn{1}{c}{$5e^{-5}$} & $1e^{-5}$ \\ \cline{2-10}
& &$\kappa$ & 0.1 & \multicolumn{1}{c}{0.1} & \multicolumn{1}{c}{0.1} & \multicolumn{1}{c}{0.1}  & 0.1\\ 
\multirow{-6}{*}{P2-R} & \multirow{-2}{*}{$20\%$} &$\xi$ & $5e^{-5}$ & \multicolumn{1}{c}{$5e^{-5}$} & \multicolumn{1}{c}{$5e^{-5}$} & \multicolumn{1}{c}{$1e^{-5}$} & $5e^{-5}$  \\ \bottomrule
\end{tabular}
\end{table}

After the complex multi-echo images $\{\vz_i\ |_{i=1}^I\}$ are recovered using the least squares, $l_1$-norm, and AMP-PE approaches, we can extract magnitude and phase images respectively from each approach to be used for QSM reconstruction. The phase images are first unwrapped using Laplacian-based phase unwrapping \cite{Laplacian:L1:2014}, and the background field is then removed using PDF \cite{PDF:Liu:2011}. The background-removed phase images are converted to produce the local field maps for each echo, and the average local field map $\vb$ is used for QSM reconstruction. The susceptibility $\chi$ is then recovered from $\vb$ using the nonlinear MEDI algorithm \cite{Liu:MEDI:2012,Liu:MEDI:2013}:
\begin{align}
    \min_{\chi}\quad\zeta\cdot\left\|\mW\big(\exp(i\mD\chi)-\exp(i\vb)\big)\right\|_2+\|\mM\nabla\chi\|_1\,,
\end{align} 
where $\mW$ is a weighting matrix that compensates for the phase noise, $\mD\chi$ performs the convolution of $\chi$ with the dipole kernel in the Fourier space, $\zeta$ is the parameter that emphasizes the data-fidelity term. Inside the $l_1$-regularization term, $\nabla$ is the spatial gradient operator on $\chi$, $\mM$ is the weighting mask that assigns zero to gradients from structural edges and assigns one to all other gradients computed from magnitude images, $\mM$ is also computed from the magnitude image. In the nonlinear MEDI algorithm, $50\%$ of pixels are selected as edges, and $\zeta$ is chosen to be $25000$.

\subsection{Evaluation Criteria}

The ground-truth reference images of $\vz_0,\vr_2^*$ and QSM are recovered from fully-sampled data using the least squares approach. Taking the recovered $\widehat{\vr}_2^*$ image from undersampled data for example, we use the following two criteria for the comparison of the three approaches:
\begin{enumerate}[label={\arabic*)}]
\item The pixel-wise absolute error (PAE) $e_i$:
\begin{align}
    \label{eq:pixelwise_rel_error}
    e_i=|{\hat{r}_{2i}^*}-{r_{2i}^*}|\,,
\end{align}
where ${\hat{r}_{2i}^*}$ is the $i$-th pixel of the recovered $\widehat{\vr}_2^*$ image, and ${r_{2i}^*}$ is the $i$-th pixel of the ground-truth reference $\vr_2^*$ image.
\item The normalized absolute error (NAE):
\begin{align}
    \textnormal{NAE}=\frac{\sum_i|{\hat{r}_{2i}^*}-{r_{2i}^*}|}{\sum_i|{r_{2i}^*}|}\,,
\end{align}
where the summation is over all the pixels in the image.
\end{enumerate}
The pixel-wise absolute error gives us a closer look at regions of interest locally, it showcases the localized error in the error map. Whereas the normalized absolute error offers a global picture about the difference between the recovered $\widehat{\vr}_2^*$ and the reference $\vr_2^*$. By combining the two criteria, we can get a more complete picture of the performances of the three approaches.

\section{Results}
\label{sec:exp}

The reconstructions of high-resolution 3D images are performed on the MATLAB platform using a machine (Intel Xeon Gold 5218 Processor, 2.30GHz) with 200 Gb RAM, where 6 CPUs are reserved for computation. The reconstruction times of different approaches depend on the sizes of datasets, and they are shown in Table \ref{tab:recon_time}. With the spatial resolution, FOV, and the number of echoes fixed, the size of dataset is determined by the undersampling rate. We can see that the least squares approach is the fastest one. For the $l_1$-norm regularization (L1) approach, it is faster to perform parameter tuning on a training set and use the optimized parameters on the test set: Table \ref{tab:recon_time} records the time for the L1 approach with parameter tuning to reconstruct images using one set of optimized parameters. Whereas the L-curve method computes empirical parameters for each test set and thus takes a much longer time to finish: Table \ref{tab:recon_time} records the total time for the L1 approach with L-curve to exhaustively search through all of the parameter values to select the best recovery. The proposed AMP-PE approach recovers the images and parameters jointly, it is faster than both variants of the $l_1$-norm regularization approach.

\begin{table}[tbp]
\caption{Reconstruction times ($\sim$hours) of different approaches with respect to different datasets.}
\vspace{0.5em}
\label{tab:recon_time}
\centering
\begin{tabular}{lccccccc}
\toprule
& & \multicolumn{2}{c}{\bfseries $l_1$-norm regularization} & \\ \cline{3-4}  
\multirow{-2}{*}{\bfseries Sampling rate} &\multirow{-2}{*}{\bfseries Least squares} & Parameter tuning & L-curve &\multirow{-2}{*}{\bfseries AMP-PE} \\ \midrule
$10\%$ & 3 & 31 & 195 & 15   \\ 
$15\%$ & 3.5 & 35 & 250 & 21  \\
$20\%$ & 4.5 & 40 & 300 & 25 \\
\bottomrule
\end{tabular}
\end{table}

\begin{table}[tbp]
\caption{Retrospective undersampling (P1-R): normalized absolute errors of recovered images.}
\label{tab:nae_z0_r2star_qsm_p1_r}
\centering
\resizebox{\columnwidth}{!}{
\begin{tabular}{llcccccccccccc}
\toprule
& & \multicolumn{4}{c}{$10\%$} &\multicolumn{4}{c}{$15\%$} &\multicolumn{4}{c}{$20\%$} \\ \cmidrule(lr){3-6} \cmidrule(lr){7-10} \cmidrule(lr){11-14}  
& & LSQ & L1-T & L1-L & AMP & LSQ & L1-T & L1-L & AMP & LSQ & L1-T & L1-L & AMP  \\ \cmidrule(lr){1-2} \cmidrule(lr){3-6} \cmidrule(lr){7-10} \cmidrule(lr){11-14} 
& S2 & 0.113 & 0.068 & 0.068 & \bf{0.056} & 0.077 & 0.045 & 0.045 & \bf{0.042} & 0.063 & 0.040 & 0.040 & \bf{0.038} \\
& S3 & 0.103 & 0.060 & 0.060 & \bf{0.050} & 0.072 & 0.040 & 0.041 & \bf{0.038} & 0.059 & 0.036 & 0.036 & \bf{0.034} \\
& S4 & 0.112 & 0.067 & 0.068 & \bf{0.056} & 0.081 & 0.045 & 0.046 & \bf{0.043} & 0.066 & 0.040 & 0.040 & \bf{0.038} \\
& S5 & 0.113 & 0.075 & 0.075 & \bf{0.066} & 0.075 & 0.052 & 0.052 & \bf{0.050} & 0.063 & \bf{0.046} & \bf{0.046} & \bf{0.046} \\
& S6 & 0.115 & 0.076 & 0.076 & \bf{0.065} & 0.078 & 0.052 & 0.052 & \bf{0.050} & 0.065 & \bf{0.046} & \bf{0.046} & \bf{0.046} \\
\multirow{-6}{*}{$\hat{\vz}_0$} & S7 & 0.109 & 0.070 & 0.070 & \bf{0.061} & 0.074 & 0.048 & 0.048 & \bf{0.047} & 0.061 & \bf{0.043} & 0.044 & \bf{0.043} \\ \cmidrule(lr){1-2} \cmidrule(lr){3-6} \cmidrule(lr){7-10} \cmidrule(lr){11-14} 
& S2 & 0.322 & 0.212 & 0.212 & \bf{0.174} & 0.221 & 0.139 & 0.141 & \bf{0.131} & 0.183 & 0.123 & 0.125 & \bf{0.118} \\
& S3 & 0.322 & 0.212 & 0.212 & \bf{0.171} & 0.224 & 0.137 & 0.139 & \bf{0.127} & 0.183 & 0.120 & 0.121 & \bf{0.114} \\
& S4 & 0.349 & 0.233 & 0.235 & \bf{0.187} & 0.251 & 0.149 & 0.151 & \bf{0.140} & 0.206 & 0.130 & 0.131 & \bf{0.125} \\
& S5 & 0.343 & 0.256 & 0.256 & \bf{0.222} & 0.226 & 0.167 & 0.167 & \bf{0.161} & 0.188 & 0.146 & 0.146 & \bf{0.145} \\
& S6 & 0.347 & 0.254 & 0.254 & \bf{0.215} & 0.232 & 0.165 & 0.165 & \bf{0.159} & 0.193 & 0.146 & 0.146 & \bf{0.144} \\
\multirow{-6}{*}{$\hat{\vr}_2^*$} & S7 & 0.340 & 0.244 & 0.244 & \bf{0.208} & 0.226 & 0.158 & 0.158 & \bf{0.152} & 0.187 & 0.140 & 0.141 & \bf{0.138} \\ \cmidrule(lr){1-2} \cmidrule(lr){3-6} \cmidrule(lr){7-10} \cmidrule(lr){11-14}  
& S2 & 0.525 & 0.395 & 0.395 & \bf{0.347} & 0.328 & 0.254 & 0.253 & \bf{0.235} & 0.268 & 0.218 & 0.218 & \bf{0.205} \\
& S3 & 0.556 & 0.416 & 0.416 & \bf{0.342} & 0.345 & 0.259 & 0.259 & \bf{0.234} & 0.274 & 0.219 & 0.219 & \bf{0.202} \\
& S4 & 0.527 & 0.401 & 0.401 & \bf{0.346} & 0.345 & 0.257 & 0.257 & \bf{0.230} & 0.278 & 0.210 & 0.210 & \bf{0.196} \\
& S5 & 0.500 & 0.407 & 0.407 & \bf{0.374} & 0.297 & 0.256 & 0.256 & \bf{0.241} & 0.238 & 0.211 & 0.211 & \bf{0.205} \\
& S6 & 0.494 & 0.386 & 0.386 & \bf{0.351} & 0.299 & 0.244 & 0.244 & \bf{0.230} & 0.241 & 0.209 & 0.209 & \bf{0.198} \\
\multirow{-6}{*}{$\hat{\chi}$} & S7 & 0.552 & 0.425 & 0.425 & \bf{0.386} & 0.330 & 0.269 & 0.269 & \bf{0.252} & 0.264 & 0.223 & 0.223 & \bf{0.217} \\
\bottomrule
\end{tabular}
}
\end{table}

\begin{table}[tbp]
\caption{Retrospective undersampling (P2-R): normalized absolute errors of recovered images.}
\label{tab:nae_z0_r2star_qsm_p2_r}
\centering
\resizebox{\columnwidth}{!}{
\begin{tabular}{llcccccccccccc}
\toprule
& & \multicolumn{4}{c}{$10\%$} &\multicolumn{4}{c}{$15\%$} &\multicolumn{4}{c}{$20\%$} \\ \cmidrule(lr){3-6} \cmidrule(lr){7-10} \cmidrule(lr){11-14}  
& & LSQ & L1-T & L1-L & AMP & LSQ & L1-T & L1-L & AMP & LSQ & L1-T & L1-L & AMP  \\ \cmidrule(lr){1-2} \cmidrule(lr){3-6} \cmidrule(lr){7-10} \cmidrule(lr){11-14} 
& S9 & 0.107 & 0.069 & 0.069 & \bf{0.061} & 0.070 & 0.047 & 0.047 & \bf{0.046} & 0.058 & \bf{0.042} & \bf{0.042} & \bf{0.042} \\
& S10 & 0.122 & 0.083 & 0.083 & \bf{0.071} & 0.080 & 0.054 & 0.054 & \bf{0.053} & 0.066 & \bf{0.048} & \bf{0.048} & \bf{0.048} \\
& S11 & 0.114 & 0.078 & 0.078 & \bf{0.065} & 0.071 & 0.050 & 0.050 & \bf{0.048} & 0.059 & \bf{0.044} & 0.045 & \bf{0.044} \\
\multirow{-4}{*}{$\hat{\vz}_0$} & S12 & 0.104 & 0.067 & 0.067 & \bf{0.058} & 0.067 & 0.046 & 0.046 & \bf{0.045} & 0.056 & \bf{0.041} & \bf{0.041} & \bf{0.041} \\ \cmidrule(lr){1-2} \cmidrule(lr){3-6} \cmidrule(lr){7-10} \cmidrule(lr){11-14} 
& S9 & 0.322 & 0.231 & 0.231 & \bf{0.196} & 0.211 & 0.148 & 0.150 & \bf{0.144} & 0.174 & \bf{0.131} & \bf{0.131} & \bf{0.131} \\
& S10 & 0.339 & 0.250 & 0.250 & \bf{0.215} & 0.223 & 0.161 & 0.161 & \bf{0.156} & 0.186 & 0.143 & 0.143 & \bf{0.142} \\
& S11 & 0.325 & 0.245 & 0.245 & \bf{0.202} & 0.206 & 0.151 & 0.152 & \bf{0.146} & 0.170 & \bf{0.133} & 0.134 & \bf{0.133} \\
\multirow{-4}{*}{$\hat{\vr}_2^*$} & S12 & 0.293 & 0.212 & 0.212 & \bf{0.179} & 0.190 & 0.137 & 0.138 & \bf{0.132} & 0.159 & 0.122 & 0.122 & \bf{0.121} \\ \cmidrule(lr){1-2} \cmidrule(lr){3-6} \cmidrule(lr){7-10} \cmidrule(lr){11-14}  
& S9 & 0.482 & 0.389 & 0.389 & \bf{0.356} & 0.289 & 0.243 & 0.243 & \bf{0.230} & 0.231 & 0.203 & 0.203 & \bf{0.197} \\
& S10 & 0.531 & 0.424 & 0.424 & \bf{0.390} & 0.312 & 0.261 & 0.261 & \bf{0.246} & 0.251 & 0.219 & 0.219 & \bf{0.211} \\
& S11 & 0.455 & 0.363 & 0.363 & \bf{0.319} & 0.264 & 0.221 & 0.221 & \bf{0.211} & 0.212 & 0.186 & 0.186 & \bf{0.183} \\
\multirow{-4}{*}{$\hat{\chi}$} & S12 & 0.472 & 0.376 & 0.376 & \bf{0.338} & 0.276 & 0.235 & 0.235 & \bf{0.222} & 0.222 & 0.197 & 0.197 & \bf{0.191} \\
\bottomrule
\end{tabular}
}
\end{table}

\subsection{Retrospective Undersampling}
Using a brain mask, we compute the pixel-wise absolute error (PAE) and normalized absolute error (NAE) with respect to the brain region. The NAEs of recovered initial magnetization $\widehat{\vz}_0$, recovered $R_2^*$ map $\widehat{\vr}_2^*$ and recovered QSM are given in Tables \ref{tab:nae_z0_r2star_qsm_p1_r} and \ref{tab:nae_z0_r2star_qsm_p2_r}. The computed PAEs are given in Tables S1 and S2 of the Supporting Information due to space limitation. We can see that the proposed AMP-PE approach performs better than the other approaches in general, except a few cases where the L1 approaches perform as well as AMP-PE. The least squares (LSQ) approach does not require parameter tuning. It simply minimizes the mean squared error of the imaging forward model, and does not use any prior information to help the reconstruction, which thus leads to the worst performance. However, the solution from the LSQ approach could serve as a valuable initialization for the other approaches. The $l_1$-norm regularization approach enforces the sparse prior on the wavelet coefficients through the $l_1$-norm regularizer. The regularization parameter is either tuned on a training set (L1-T) or estimated using the heuristic L-curve method (L1-L). Apart from the sparse prior on wavelet coefficients, the proposed AMP-PE approach also incorporates additional information from the mono-exponential decay model. This allows AMP-PE to achieve better performance than the L1 approach. AMP-PE treats the distribution parameters as unknown variables, it automatically and adaptively estimates them with respect to each dataset.

\begin{figure}[tbp]
\centering
\includegraphics[width=\textwidth]{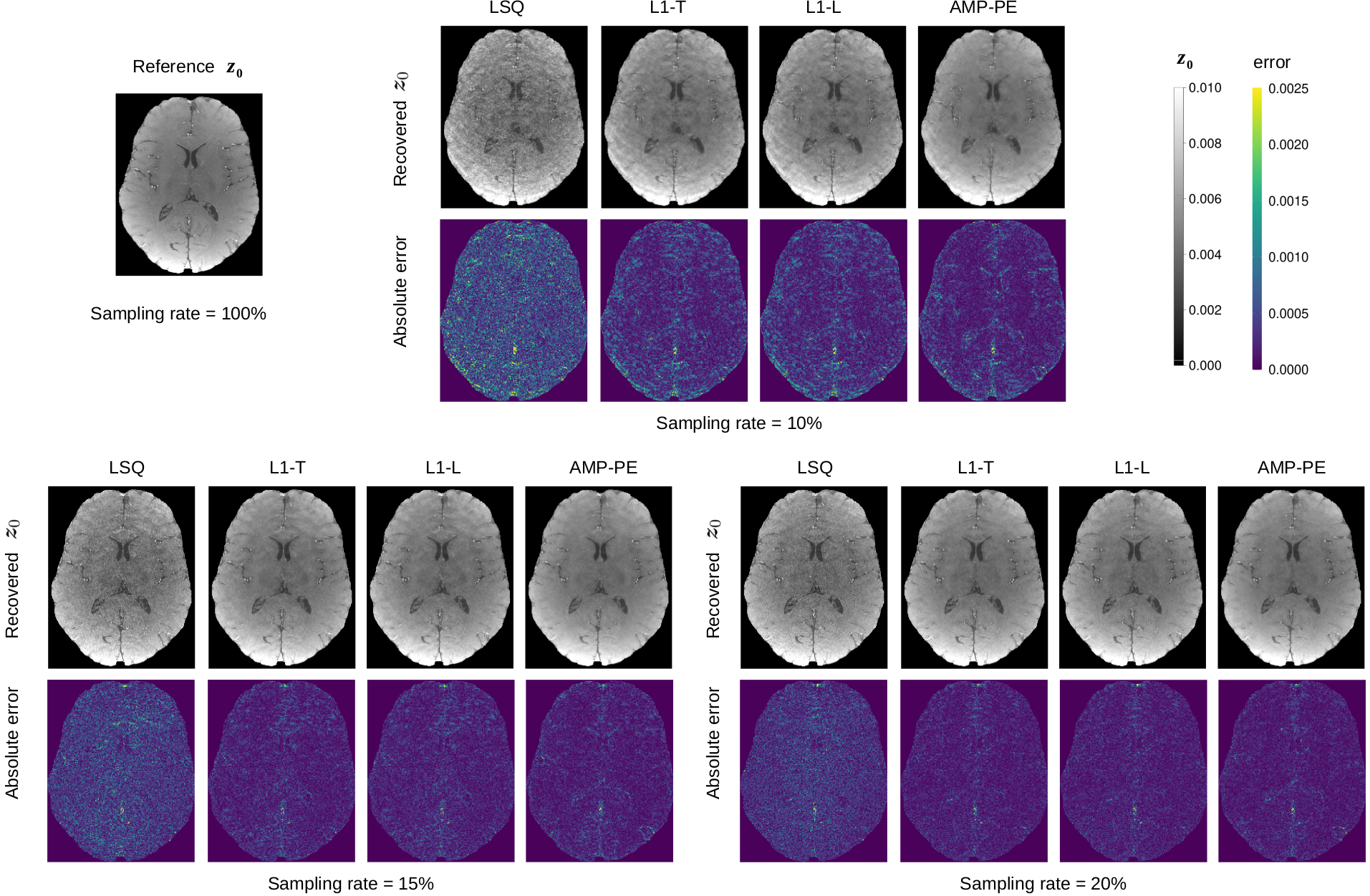}
\caption{Retrospective undersampling: recovered initial magnetization $\widehat{\vz}_0$ using the least squares approach (LSQ), the $l_1$-norm regularization approach with parameter tuning (L1-T) and L-curve method (L1-L), the proposed AMP-PE approach.} 
\label{fig:rel_error_x0}
\end{figure}

\begin{figure}[tbp]
\centering
\includegraphics[width=\textwidth]{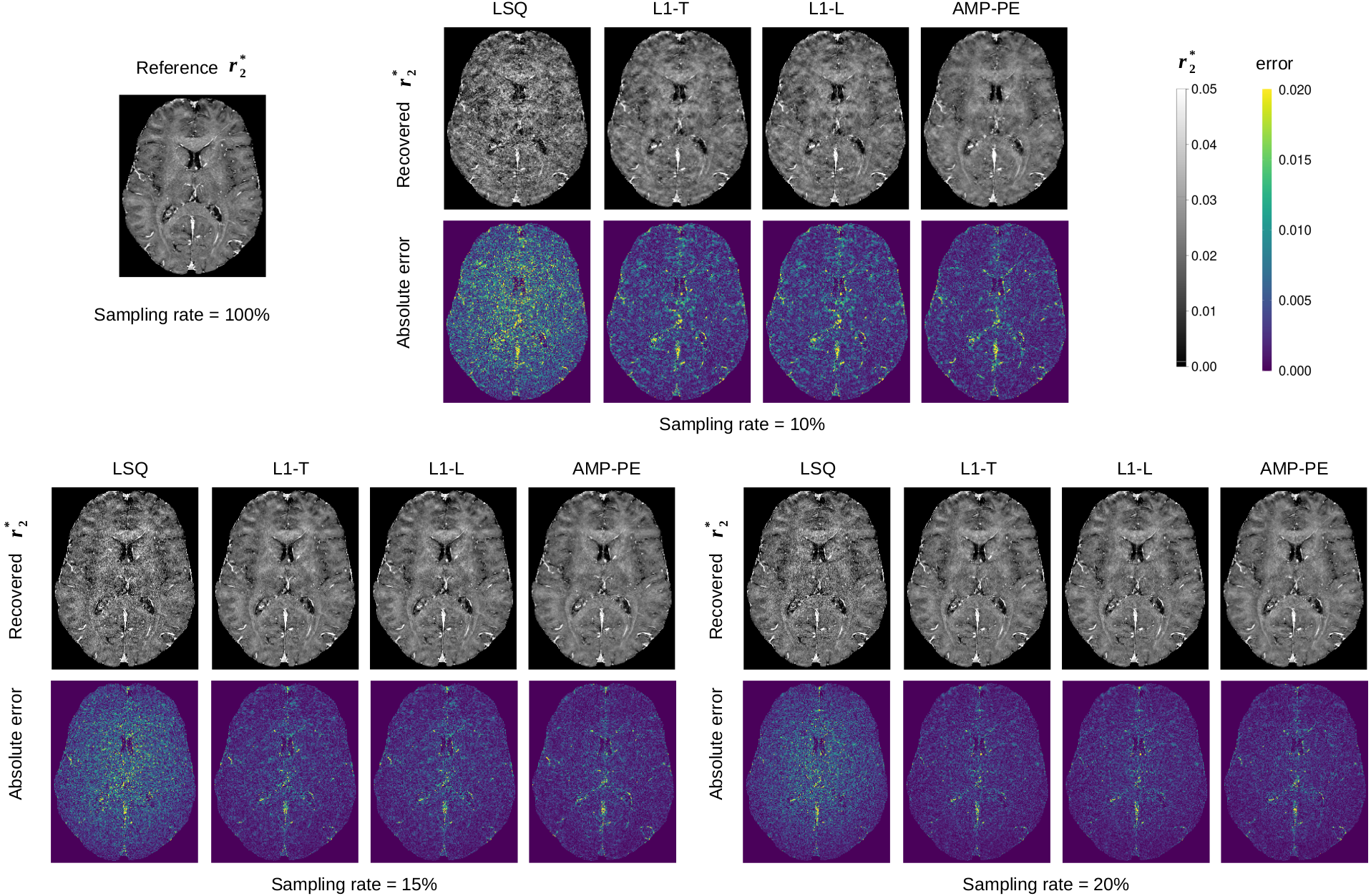}
\caption{Retrospective undersampling: recovered $R_2^*$ map $\widehat{\vr}_2^*$ using the least squares approach (LSQ), the $l_1$-norm regularization approach with parameter tuning (L1-T) and L-curve method (L1-L), the proposed AMP-PE approach.} 
\label{fig:rel_error_r2star}
\end{figure}

\begin{figure}[tbp]
\centering
\includegraphics[width=\textwidth]{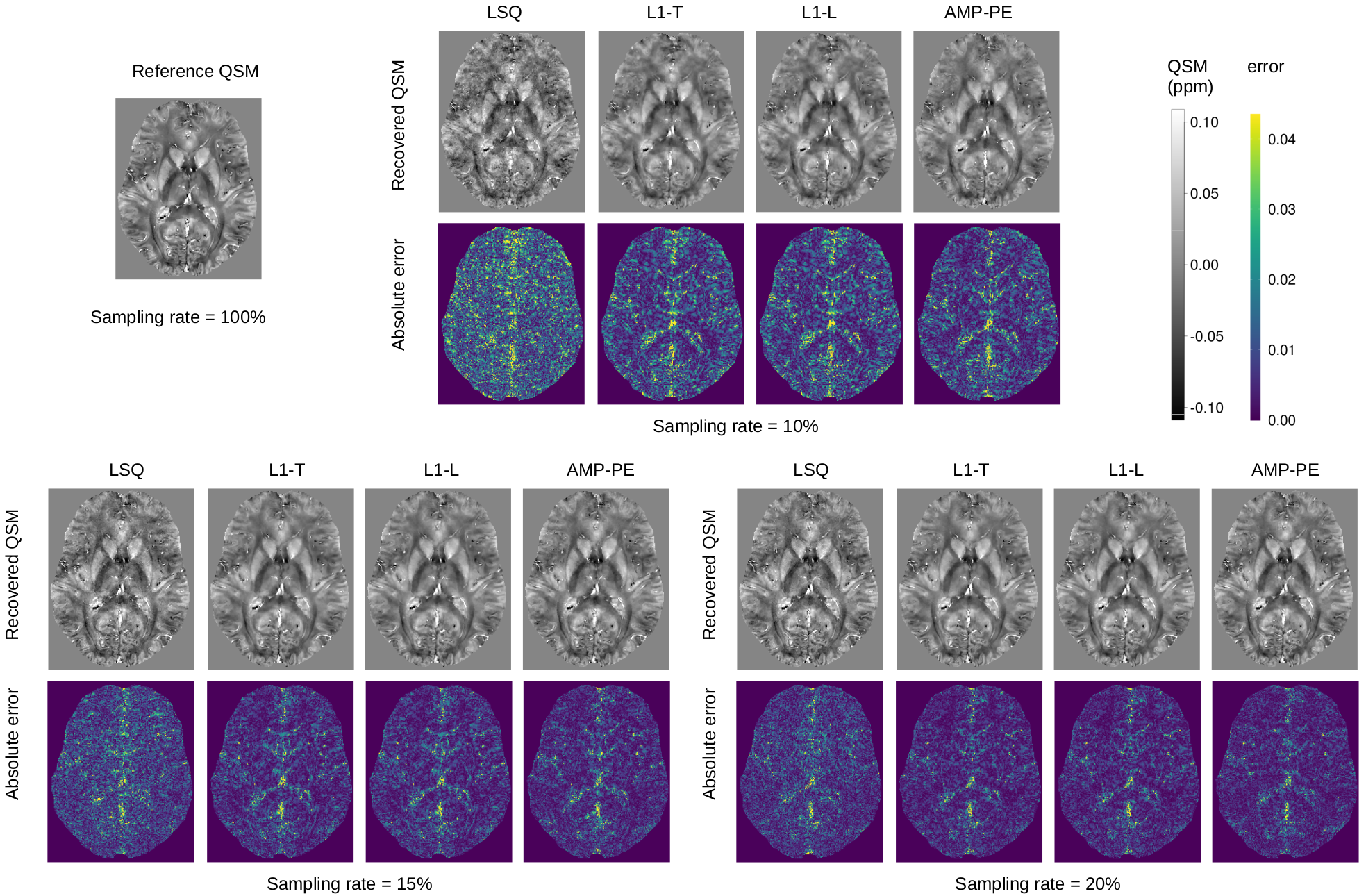}
\caption{Retrospective undersampling: recovered QSM using the least squares approach (LSQ), the $l_1$-norm regularization approach with parameter tuning (L1-T) and L-curve method (L1-L), the proposed AMP-PE approach.} 
\label{fig:rel_error_qsm}
\end{figure}

Taking one slice from the recovered 3D brain image from ``S2'' for example, we show the recovered images and the errors in Fig. \ref{fig:rel_error_x0}-Fig. \ref{fig:rel_error_qsm}. In particular, Fig. \ref{fig:rel_error_qsm} shows the axial view of recovered QSM. In order to assess the streaking artifacts in QSM, additional coronal and sagittal views are provided in Figures S1 and S2 of the Supporting Information. When the sampling rate is $10\%$, we can see that the images recovered by the least squares and the $l_1$-norm regularization approaches are noisier compared to those recovered by the AMP-PE approach. When the sampling rate further increases to $15\%$ and $20\%$, the proposed AMP-PE approach still leads in image quality, while the differences between the three approaches become smaller.


\subsection{Prospective Undersampling}
We then compare the recovery approaches on the datasets acquired using two prospective protocols, where the undersampling rates vary in $\{10\%,\ 15\%,\ 20\%,\ 100\%\}$. In this case, the images recovered from fully-sampled datasets serve as the reference images as before. However, since the undersampled datasets were acquired independently from the fully-sampled dataset, the noise profiles in these datasets were also independent and different. The variation of noise leads to a bias in the recovered reference image from fully-sampled data. As a result, the errors with respect to the reference image are also much larger compared to the retrospective case due to the lack of a ``ground-truth'' image in the prospective case. 

\begin{figure}[tbp]
\centering
\subfigure{
\includegraphics[width=\textwidth]{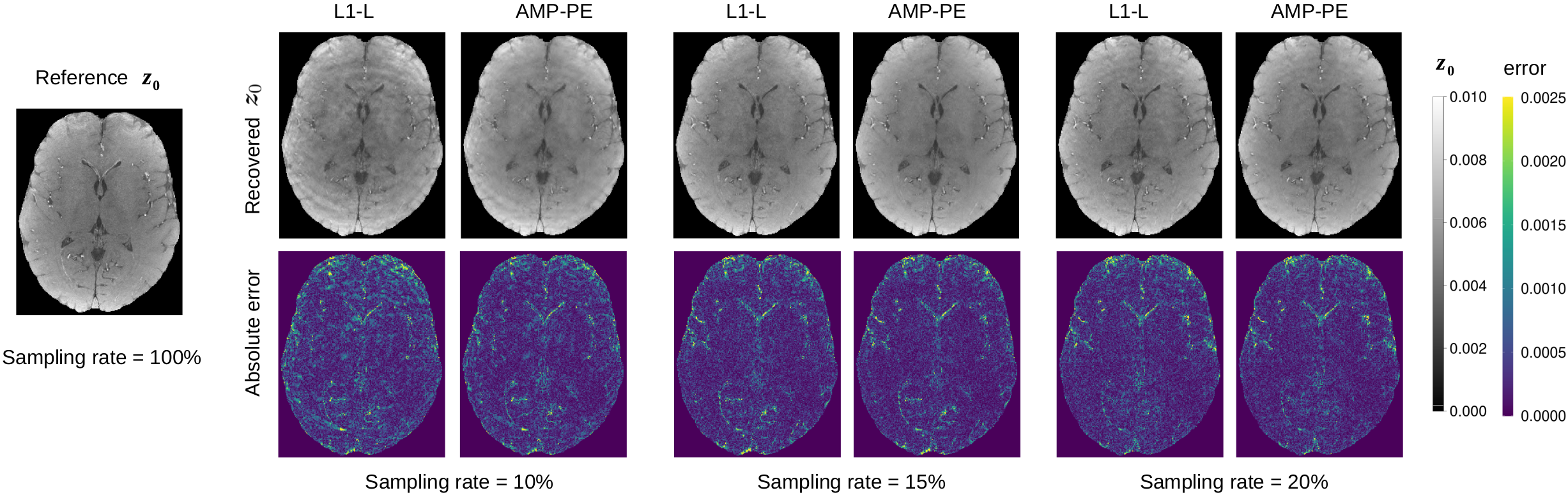}}
\subfigure{
\includegraphics[width=\textwidth]{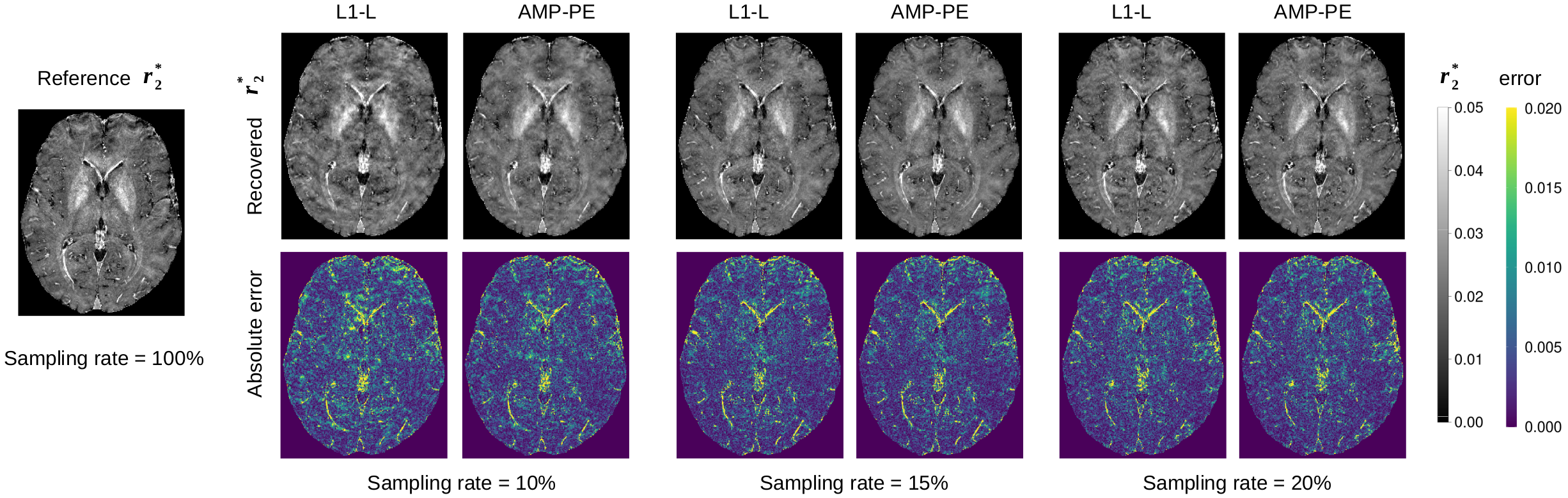}}
\subfigure{
\includegraphics[width=\textwidth]{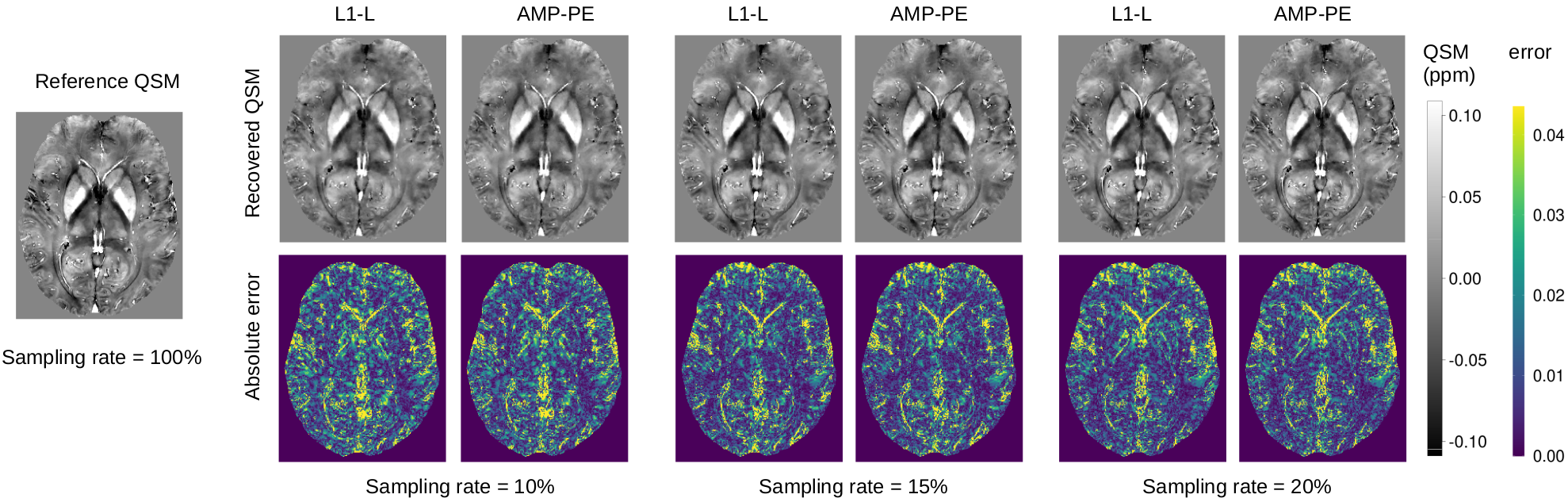}}
\caption{Prospective undersampling: recovered initial magnetization $\hat{\vz_0}$, $R_2^*$ map $\hat{\vr}_2^*$ and QSM using the $l_1$-norm regularization approach with L-curve method (L1-L) and the proposed AMP-PE approach.} 
\label{fig:rel_error_x0_r2star_qsm_prospective}
\end{figure}

The normalized absolute errors (NAE) and pixel-wise absolute errors (PAE) contain biases and are given in Tables S6-S9 of the Supporting Information due to space limitation. We should note that bias of the reference image makes the computed errors larger, as compared to the corresponding retrospective case with the same fully-sampled dataset. Based on the biased quantitative results, it is hard to evaluate different approaches, and they are thus for reference purposes only. Taking one slice from the recovered 3D brain image S9 for example, we also show the recovered images and their errors from the L1 approach with the L-curve method (L1-L) and the AMP-PE approach in Fig. \ref{fig:rel_error_x0_r2star_qsm_prospective}. We can see that the prospective undersampling scheme does produce comparable and consistent results to the retrospective case through visual inspection.

\section{Discussion}

We use undersampling to reduce the scan time required for high-resolution 3D imaging, and rely on compressive sensing (CS) to fill in the missing information. It has been shown in CS theory that the more incoherent the measurement operator is, the better the recovery performance can be \cite{RUP06,CS06}. Random sampling has been widely used in CS to construct such an incoherent operator. However, when we are sampling in the $k$-space, the Poisson-disk sampling is a better choice: it keeps the randomness while imposing a minimum-distance constraint between any two sampling locations. As shown in Fig. S11 of the Supporting Information, the sampling locations are thus more uniformly spread across the $k$-space compared to random sampling, leading to a group of diverse measurement vectors. We compared the performances of random sampling and Poisson-disk sampling in Figures S12-S14 of the Supporting Information. The results show that Poisson-disk sampling is better at removing aliasing artifacts from the images and produces lower errors than random sampling.

The L1 approach requires suitable parameters to recover the images successfully. From Table \ref{tab:parameter_1st_protocol}, we can see that the tuned parameters and the parameters determined by the L-curve method are close to or the same as each other. The optimal parameters obtained through an exhaustive search on the test set are given in Table S10 of the Supporting Information, and they are also close to or the same as Table \ref{tab:parameter_1st_protocol}. The reason why the working parameters in the L1 approach are stable can be explained from a Bayesian perspective. Take the problem in \eqref{eq:l1_norm_multi_echo} for example, when Laplace distribution is chosen as the signal prior $p(\vv|\lambda)$ in \eqref{eq:laplace_dist} and additive white Gaussian distribution is chosen as the noise prior $p(\vy|\vv,\theta)$ in \eqref{eq:awgn}, the MAP estimation of $\vv$ in \eqref{eq:map_v_est} is equivalent to the $l_1$-norm minimization problem \eqref{eq:l1_norm_multi_echo} as derived in Section S-III-B of the Supporting Information. We then have the optimal regularization parameter $\kappa=2\lambda\theta^2$. We can see that as long as the type of signal (that determines $\lambda$) and the noise level (that determines $\theta$) remain generally constant, the optimal parameter $\kappa$ should be stable and robust across different subjects.

Choosing a proper prior distribution for the wavelet coefficients $\vv$ is important for AMP-PE to achieve a successful recovery. We used the Laplace distribution given in \eqref{eq:laplace_dist} in this paper. Another popular distribution for modelling sparse signals is the Bernoulli-Gaussian mixture (BGM) distribution. The results obtained from AMP-PE using the two distributions are compared in Fig. S15 of the Supporting Information. We can see that the Laplace prior performs better than the BGM prior. As shown in Fig. S15, the image recovered with BGM prior is oversmoothed and has higher errors. In practice, the fitness of a distribution can be measured by the log-likelihood of the coefficients $\vv$ under such distribution. Using the ground-truth wavelet coefficients $\vv$ of the image in Fig. S15 as the data, we then compute the log-likelihoods of $\vv$ under the two priors. The log-likelihood from Laplace prior is $5.36e^5$ and the log-likelihood from BGM prior is $3.91e^5$, indicating that Laplace prior is a better fit in this case.

We can see from Fig. \ref{fig:rel_error_qsm} that there is an evident loss in sharpness and contrast in fine structures of the recovered QSM. In order to further investigate this, we showed the local field maps produced from phase images in Figures S3-S5 of the Supporting Information. We can see that the loss of high-frequency structures already occurred in the local field maps, and was carried over to subsequent QSM. The loss is caused by the combined effect of undersampling and regularization. First, to assess the effect of undersampling, let's look at the recovered images from the LSQ approach that minimizes the data-fidelity term alone and does not use regularization. When the sampling rate is low (say $10\%$), the acquired k-space measurements do not contain enough high-frequency data that contributes to details in the recovered image. As the sampling rate is increased, more high-frequency data are incorporated to the measurements, and more details start to emerge in the images from LSQ. Second, when the L1 and AMP-PE approach use regularization (sparse prior) to improve the image quality, they set the low-energy wavelet coefficients to zero. The cut-off threshold of wavelet coefficients is determined on a global scale, and this will inevitably wash out some details further from the image. The key is thus to balance the trade-off between data-fidelity and regularization via parameter tuning or estimation.

The AMP approach has been shown to be more computationally efficient than the L1 approach \cite{Vila:EMGM:2013}. Table \ref{tab:recon_time} also shows that the AMP-PE approach is faster than the L1 approach. However, due to the large size of 3D datasets, compressive sensing methods like the L1 and AMP approaches still require long computational times. Since undersampling takes place along the phase-encoding directions and the readout direction is fully sampled, one solution is to perform FFT along the readout direction and decompose the 3D reconstruction into parallelizable 2D reconstructions, though the overall performance would drop a bit due to the switch from 3D wavelet basis to 2D wavelet basis. If GPU is available, a better option would be to take advantage of GPU computing in MATLAB to speed up the 3D recovery.

\section{Conclusion}
\label{sec:con}
In order to improve the quality of $R_2^*$ map and QSM recovered from undersampled data and to avoid manual parameter tuning, we propose a Bayesian approach to combine a mono-exponential decay model with a sparse prior on the wavelet coefficients of images. In particular, the wavelet coefficients are assumed to be independently generated from a sparsity-promoting distribution, and the measurement noise is assumed to be additive white Gaussian noise. The incorporation of mono-exponential decay model allows us to achieve better performance than the state-of-the-art $l_1$-norm regularization approach that only uses sparse prior information. By treating the distribution parameters as unknown variables \cite{PE_GAMP17}, we can jointly recover the parameters with the wavelet coefficients of images under the proposed nonlinear-AMP framework. Compared to other compressive sensing methods that enforce the sparse prior through regularization, our proposed approach does not require manual parameter tuning: the distribution parameters are automatically and adaptively estimated with respect to each dataset. It thus could work with a clinical, prospective undersampling scheme where parameter tuning is often impossible or difficult due to the lack of ground-truth image.

\pagebreak

\begin{appendices}
\renewcommand\thetable{\thesection\arabic{table}}
\renewcommand\thefigure{\thesection\arabic{figure}}

\section{Messages Exchanged between the Variable and Factor Nodes}

\subsection{Recovery of Multi-echo Image Distribution}
\label{subsec:messages_multi_echo_image_distribution}
In the following we derive the messages exchanged on the factor graph in Fig. \ref{fig:factor_graph_multi_echo}, which are used to recovery the multi-echo image distribution $p_{\mathcal{M}}(z_{in}|\vy)$ in \eqref{eq:multi_echo_image_prior}.
\begin{itemize}
    \item Specifically, we have the following messages passed from $\Phi_{im}$ to $\lambda_i$ in the $(t+1)$-th iteration.
    \begin{subequations}
    \label{eq:sp_fv_signal}
    \begin{align}
    \label{eq:sp_fv_phi_x}
    \begin{split}
    &\Delta^{(t+1)}_{\Phi_{im}\rightarrow v_{in}}=C+\log\int_{\vv_i\backslash v_{in}}\Phi\left(y_{im}, \vv_i, \hat{\theta}^{(t)}_{\mathcal{M}}\right)\cdot\exp\Big(\sum_{l\neq n}\Delta^{(t)}_{v_{il}\rightarrow\Phi_{im}}\Big)
    \end{split}\\
    \label{eq:sp_vf_z_omega}
    &\Delta^{(t+1)}_{v_{in}\rightarrow \Omega_{in}}=\sum_k\Delta^{(t+1)}_{\Phi_{ik}\rightarrow v_{in}}\\
    \label{eq:sp_fv_omega_lambda}
    &\Delta^{(t+1)}_{\Omega_{in}\rightarrow\lambda_i}=C+\log\int_{v_{in}}\Omega(v_{in},\lambda_i)\cdot\exp\Big(\Delta^{(t+1)}_{v_{in}\rightarrow \Omega_{in}}\Big)\,,
    \end{align}
    \end{subequations}
    where $C$ (by abuse of notation\footnote{Note that the $C$ in \eqref{eq:sp_fv_phi_x} and the $C$ in \eqref{eq:sp_fv_omega_lambda} are different, they are both some constants in the $(t+1)$-th iteration. To simplify the notations, $C$ is reserved to denote the constant in the rest of messages as well.}) denotes a constant that depends on variables in the previous $t$-th iteration, $\vv_i\backslash v_{in}$ is the vector $\vv_i$ with its $n$-th entry $v_{in}$ removed. The sparse signal prior distribution $\Omega(v_{in},\lambda_i)= p(v_{in}|\lambda_i)$ is given in \eqref{eq:laplace_dist}. Let $\vf_i=\mA_i\mH^{-1}\vv_i$ denote the noiseless measurement in the $i$-th echo, where $\mH^{-1}$ is the inverse wavelet transform matrix. The noisy measurement in the $i$-th echo is $\vy_i$, and the total measurement $\vy=[\vy_1^T\ \cdots\ \vy_I^T]^T$. Under the AWGN model given in \eqref{eq:awgn}, the noisy measurement distribution $\Phi(y_{im}, \vv_i,\theta_{\mathcal{M}})=p(y_{im}|f_{im},\theta^2)=\mathcal{N}(y_{im}|f_{im},\theta^2_{\mathcal{M}})$. 
    
    \item We further have the following messages passed from $\Omega_{in}$ to $\theta_{\mathcal{M}}$ in the $(t+1)$-th iteration:
    \begin{subequations}
    \begin{align}
    \label{eq:sp_fv_omega_x}
    &\Delta^{(t+1)}_{\Omega_{in}\rightarrow v_{in}}=C+\log\Omega\Big(v_{in},\hat{\lambda}_i^{(t+1)}\Big)\\
    \label{eq:sp_vf_z_phi}
    &\Delta^{(t+1)}_{v_{in}\rightarrow \Phi_{im}}=\Delta^{(t+1)}_{\Omega_{in}\rightarrow v_{in}}+\sum_{k\neq m}\Delta^{(t+1)}_{\Phi_{ik}\rightarrow v_{in}}\\
    \label{eq:sp_fv_phi_theta}
    &\Delta^{(t+1)}_{\Phi_{im}\rightarrow\theta_{\mathcal{M}}}=C+\log\int_{\vv_i}\Phi(y_{im},\vv_i,\theta_{\mathcal{M}})\cdot\exp\Big(\sum_l\Delta^{(t+1)}_{v_{il}\rightarrow\Phi_{im}}\Big)\,.
    \end{align}
    \end{subequations}
\end{itemize}

\subsection{Recovery of $R_2^*$ Map}
\label{subsec:messages_t2star_map}
In the following we derive the messages exchanged on the factor graph in Fig. \ref{fig:factor_graph_r2star_multi_echo}, which recovers the $R_2^*$ map by combining the mono-exponential decay model with the multi-echo image distribution.
\begin{itemize}
    \item In the $(t+1)$-th iteration, the messages passed from $\Phi_{im}$ to $\Psi_{in}$ are
    \begin{subequations}
    \begin{align}
        &\Delta^{(t+1)}_{\Phi_{im}\rightarrow z_{in}}=C+\log\int_{\vz_i\backslash z_{in}}\Phi\left(y_{im}, \vz_i, \hat{\theta}^{(t)}_{\mathcal{E}}\right)\cdot\exp\Big(\sum_{l\neq n}\Delta^{(t)}_{z_{il}\rightarrow\Phi_{im}}\Big)\\
        &\Delta^{(t+1)}_{z_{in}\rightarrow \Gamma_{in}}=\log\Xi(z_{in})+\sum_k\Delta^{(t+1)}_{\Phi_{ik}\rightarrow z_{in}}\\
        \label{eq:Delta_to_sin}
        &\Delta^{(t+1)}_{\Gamma_{in}\rightarrow s_{in}}=C+\log\int_{z_{in}}\Gamma(s_{in},z_{in})\cdot\exp\Big(\Delta^{(t+1)}_{z_{in}\rightarrow \Gamma_{in}}\Big)\\
        &\Delta^{(t+1)}_{s_{in}\rightarrow\Psi_{in}}=\Delta^{(t+1)}_{\Gamma_{in}\rightarrow s_{in}}\,,
    \end{align}
    \end{subequations}
    where the factor node $\Gamma_{in}$ enforces the equality constraint $s_{in}=|z_{in}|$:
    \begin{align}
    \label{eq:gamma_constraint}
        \Gamma(s_{in},z_{in}) = \delta(s_{in}=|z_{in}|)\,,
    \end{align}
    where $\delta(\cdot)$ is the Dirac delta function.
    
    The messages from $\Psi_{in}$ to $\lambda_0$ are then
    \begin{subequations}
    \begin{align}
        &\Delta^{(t+1)}_{\Psi_{in}\rightarrow v_{0d}}=C+\log\int_{s_{in},\vv_0\backslash v_{0d}}\Big[\Psi(s_{in},\vv_0)\cdot\exp\Big(\Delta^{(t+1)}_{s_{in}\rightarrow\Psi_{in}}\Big)\Big]\cdot\exp\Big(\sum_{h\neq d}\Delta^{(t)}_{v_{0h}\rightarrow\Psi_{in}}\Big)\\
        &\Delta^{(t+1)}_{v_{0d}\rightarrow\Omega_{0d}}=\sum_{i,n}\Delta^{(t+1)}_{\Psi_{in}\rightarrow v_{0d}}\\
        &\Delta^{(t+1)}_{\Omega_{0d}\rightarrow\lambda_0}=C+\log\int_{v_{0d}}\Omega(v_{0d},\lambda_0)\cdot\exp\Big(\Delta^{(t+1)}_{v_{0d}\rightarrow\Omega_{0d}}\Big)\,,
    \end{align}
    \end{subequations}
    where the factor node $\Psi_{in}$ enforces the equality constraint $s_{in}=z_{0n}\cdot\exp\Big(-t_i\cdot {r_2^*}^{(t)}_n\Big)$:
    \begin{align}
    \label{eq:psi_constraint}
        \Psi(s_{in},\vv_0)=\delta\left(s_{in}=z_{0n}\cdot\exp\Big(-t_i\cdot {r_2^*}^{(t)}_n\Big)\right)\,,
    \end{align}
    where $\vz_0 = \mH^{-1}\vv_0$. By combining the above \eqref{eq:gamma_constraint} and \eqref{eq:psi_constraint}, we have encoded the mono-exponential decay model in the factor nodes $\Gamma_{in}$ and $\Psi_{in}$.
    
    \item The messages from $\Omega_{0d}$ to $s_{in}$ are
    \begin{subequations}
    \begin{align}
        &\Delta^{(t+1)}_{\Omega_{0d}\rightarrow v_{0d}}=C+\log\Omega\Big(v_{0d},\hat{\lambda}_0^{(t+1)}\Big)\\
        &\Delta^{(t+1)}_{v_{0d}\rightarrow\Psi_{in}}=\Delta^{(t+1)}_{\Omega_{0d}\rightarrow v_{0d}}+\sum_{(r,s)\neq (i,n)}\Delta^{(t+1)}_{\Psi_{rs}\rightarrow v_{0d}}\\
        \label{eq:delta_psi_s}
        &\Delta^{(t+1)}_{\Psi_{in}\rightarrow
        s_{in}}=C+\log\int_{\vv_0}\Psi(s_{in},\vv_0)\cdot\exp\Big(\sum_{h}\Delta^{(t+1)}_{v_{0h}\rightarrow\Psi_{in}}\Big)\,.
    \end{align}
    \end{subequations}
    The model parameters $\vr_2^*$ can be computed by minimizing the least square error of the mono-exponential decay model:
    \begin{align}
    \label{eq:recover_r2star}
        \hat{r}_{2n}^*=\arg\min_{{r_2^*}_n}\ \sum_{i=1}^I\Big(\hat{s}_{in}-\hat{z}_{0n}\cdot\exp\big(-t_i\cdot {r_2^*}_n\big)\Big)^2\,,
    \end{align}
    where $\hat{\vs}_{i}$ is the magnitude of the multi-echo image calculated using the message from \eqref{eq:Delta_to_sin}, and $\hat{\vz}_0$ is the recovered initial magnetization image. They can be computed via the MAP estimation
    \begin{align}
    \hat{s}_{in}&=\arg\max_{s_{in}}\ \exp\Big(\Delta^{(t+1)}_{\Gamma_{in}\rightarrow s_{in}}\Big)\\
    \hat{v}_{0n}&=\arg\max_{v_{0n}}\ p(v_{0n}|\vy)\ =\arg\max_{v_{0n}}\ \exp\Big(\Delta^{(t+1)}_{\Omega_{0n}\rightarrow v_{0n}}+\sum_{il}\Delta^{(t+1)}_{\Psi_{il}\rightarrow v_{0n}}\Big)\\
    \label{eq:recover_proton}
    \widehat{\vz}_0&=\mH^{-1}\hat{\vv}_0\,.
    \end{align}
    
    We further have the following messages passed from $s_{in}$ to $\theta_{\mathcal{E}}$:
    \begin{subequations}
    \begin{align}
        &\Delta^{(t+1)}_{s_{in}\rightarrow\Gamma_{in}}=\Delta^{(t+1)}_{\Psi_{in}\rightarrow s_{in}}\\
        &\Delta^{(t+1)}_{\Gamma_{in}\rightarrow z_{in}} = C+\log\int_{s_{in}}\Gamma(s_{in},z_{in})\cdot\exp\Big(\Delta^{(t+1)}_{s_{in}\rightarrow\Gamma_{in}}\Big)\\
        &\Delta^{(t+1)}_{z_{in}\rightarrow\Phi_{im}}=\Delta^{(t+1)}_{\Gamma_{in}\rightarrow z_{in}}+\log\Xi(z_{in})+\sum_{k\neq m}\Delta^{(t+1)}_{\Phi_{ik}\rightarrow z_{in}}\\
        &\Delta^{(t+1)}_{\Phi_{im}\rightarrow\theta_{\mathcal{E}}}=C+\log\int_{\vz_i}\Phi(y_{im},\vz_i,\theta_{\mathcal{E}})\cdot\exp\Big(\sum_l\Delta^{(t+1)}_{z_{il}\rightarrow\Phi_{im}}\Big)\,.
    \end{align}
    \end{subequations}

\end{itemize}

\end{appendices}

\bibliography{ref}

\begin{thebibliography}{10}

\bibitem{Bernstein:2004}
M.~A. Bernstein, K.~F. King, and X.~J. Zhou, {\em Handbook of MRI Pulse
  Sequences}.
\newblock Burlington, MA: Elsevier Academic Press, 2004.

\bibitem{Mamisch:T2Star:2012}
T.~C. Mamisch, T.~Hughes, T.~J. Mosher, C.~Mueller, S.~Trattnig, C.~Boesch, and
  G.~H. Welsch, ``T2 star relaxation times for assessment of articular
  cartilage at 3 t: a feasibility study,'' {\em Skeletal Radiology}, vol.~41,
  no.~3, pp.~287--292, 2012.

\bibitem{Wang:QSM:2015}
Y.~Wang and T.~Liu, ``Quantitative susceptibility mapping (qsm): Decoding mri
  data for a tissue magnetic biomarker,'' {\em Magnetic Resonance in Medicine},
  vol.~73, no.~1, pp.~82--101, 2015.

\bibitem{Langkammer:QSM:2013}
C.~Langkammer, T.~Liu, M.~Khalil, C.~Enzinger, M.~Jehna, S.~Fuchs, F.~Fazekas,
  Y.~Wang, and S.~Ropele, ``Quantitative susceptibility mapping in multiple
  sclerosis,'' {\em Radiology}, vol.~267, no.~2, pp.~551--559, 2013.

\bibitem{Deistung:QSM_R2Star:2013}
A.~Deistung, A.~Schäfer, F.~Schweser, U.~Biedermann, R.~Turner, and J.~R.
  Reichenbach, ``Toward in vivo histology: A comparison of quantitative
  susceptibility mapping (qsm) with magnitude-, phase-, and r2⁎-imaging at
  ultra-high magnetic field strength,'' {\em NeuroImage}, vol.~65,
  pp.~299--314, 2013.

\bibitem{Barbosa:QSM_R2Star:2015}
J.~H.~O. Barbosa, A.~C. Santos, V.~Tumas, M.~Liu, W.~Zheng, E.~M. Haacke, and
  C.~E.~G. Salmon, ``Quantifying brain iron deposition in patients with
  parkinson's disease using quantitative susceptibility mapping, r2 and r2*,''
  {\em Magnetic Resonance Imaging}, vol.~33, no.~5, pp.~559--565, 2015.

\bibitem{Betts:QSM_R2Star:2016}
M.~J. Betts, J.~Acosta-Cabronero, A.~Cardenas-Blanco, P.~J. Nestor, and
  E.~Düzel, ``High-resolution characterisation of the aging brain using
  simultaneous quantitative susceptibility mapping (qsm) and r2* measurements
  at 7t,'' {\em NeuroImage}, vol.~138, pp.~43--63, 2016.

\bibitem{Qiu1085}
D.~Qiu, G.-F. Chan, J.~Chu, Q.~Chan, S.-Y. Ha, M.~Moseley, and P.-L. Khong,
  ``Mr quantitative susceptibility imaging for the evaluation of iron loading
  in the brains of patients with $\beta$-thalassemia major,'' {\em American
  Journal of Neuroradiology}, vol.~35, no.~6, pp.~1085--1090, 2014.

\bibitem{Ordidge:1994}
R.~J. Ordidge, J.~M. Gorell, J.~C. Deniau, R.~A. Knight, and J.~A. Helpern,
  ``Assessment of relative brain iron concentrations using t2-weighted and
  t2*-weighted mri at 3 tesla,'' {\em Magnetic Resonance in Medicine}, vol.~32,
  no.~3, pp.~335--341, 1994.

\bibitem{McNeill:2008}
A.~McNeill, D.~Birchall, S.~J. Hayflick, A.~Gregory, J.~F. Schenk, E.~A.
  Zimmerman, H.~Shang, H.~Miyajima, and P.~F. Chinnery, ``T2* and fse mri
  distinguishes four subtypes of neurodegeneration with brain iron
  accumulation,'' {\em Neurology}, vol.~70, no.~18, pp.~1614--1619, 2008.

\bibitem{Langkammer:QSM_iron:2012}
C.~Langkammer, F.~Schweser, N.~Krebs, A.~Deistung, W.~Goessler, E.~Scheurer,
  K.~Sommer, G.~Reishofer, K.~Yen, F.~Fazekas, S.~Ropele, and J.~R.
  Reichenbach, ``Quantitative susceptibility mapping (qsm) as a means to
  measure brain iron? a post mortem validation study,'' {\em NeuroImage},
  vol.~62, no.~3, pp.~1593--1599, 2012.

\bibitem{Schweser:QSM:2012}
F.~Schweser, K.~Sommer, A.~Deistung, and J.~R. Reichenbach, ``Quantitative
  susceptibility mapping for investigating subtle susceptibility variations in
  the human brain,'' {\em NeuroImage}, vol.~62, no.~3, pp.~2083--2100, 2012.

\bibitem{Li:QSM:2011}
W.~Li, B.~Wu, and C.~Liu, ``Quantitative susceptibility mapping of human brain
  reflects spatial variation in tissue composition,'' {\em NeuroImage},
  vol.~55, no.~4, pp.~1645--1656, 2011.

\bibitem{Fazekas:1999}
F.~Fazekas, R.~Kleinert, G.~Roob, G.~Kleinert, P.~Kapeller, R.~Schmidt, and
  H.-P. Hartung, ``Histopathologic analysis of foci of signal loss on
  gradient-echo t2*-weighted mr images in patients with spontaneous
  intracerebral hemorrhage: Evidence of microangiopathy-related microbleeds,''
  {\em American Journal of Neuroradiology}, vol.~20, no.~4, pp.~637--642, 1999.

\bibitem{Kinoshita:2000}
T.~Kinoshita, T.~Okudera, H.~Tamura, T.~Ogawa, and J.~Hatazawa, ``Assessment of
  lacunar hemorrhage associated with hypertensive stroke by echo-planar
  gradient-echo t2*-weighted mri,'' {\em Stroke}, vol.~31, no.~7,
  pp.~1646--1650, 2000.

\bibitem{ORegan:2009}
D.~P. O'Regan, R.~Ahmed, N.~Karunanithy, C.~Neuwirth, Y.~Tan, G.~Durighel,
  J.~V. Hajnal, I.~Nadra, S.~J. Corbett, and S.~A. Cook, ``Reperfusion
  hemorrhage following acute myocardial infarction: Assessment with t2* mapping
  and effect on measuring the area at risk,'' {\em Radiology}, vol.~250, no.~3,
  pp.~916--922, 2009.

\bibitem{Zhang:QSM_hemorrhage:2018}
Y.~Zhang, H.~Wei, Y.~Sun, M.~J. Cronin, N.~He, J.~Xu, Y.~Zhou, and C.~Liu,
  ``Quantitative susceptibility mapping (qsm) as a means to monitor cerebral
  hematoma treatment,'' {\em Journal of Magnetic Resonance Imaging}, vol.~48,
  no.~4, pp.~907--915, 2018.

\bibitem{Sun:QSM_hemorrhage:2018}
H.~Sun, A.~C. Klahr, M.~Kate, L.~C. Gioia, D.~J. Emery, K.~S. Butcher, and
  A.~H. Wilman, ``Quantitative susceptibility mapping for following
  intracranial hemorrhage,'' {\em Radiology}, vol.~288, no.~3, pp.~830--839,
  2018.

\bibitem{Yamada:1996}
N.~Yamada, S.~Imakita, T.~Sakuma, and M.~Takamiya, ``Intracranial calcification
  on gradient-echo phase image: depiction of diamagnetic susceptibility.,''
  {\em Radiology}, vol.~198, no.~1, pp.~171--178, 1996.

\bibitem{Gupta:2001}
R.~Gupta, S.~Rao, R.~Jain, L.~Pal, R.~Kumar, S.~Venkatesh, and R.~Rathore,
  ``Differentiation of calcification from chronic hemorrhage with corrected
  gradient echo phase imaging,'' {\em Journal of Computer Assisted Tomography},
  vol.~25, pp.~698--704, Sept. 2001.

\bibitem{Deistung:QSM_calcification:2013}
A.~Deistung, F.~Schweser, B.~Wiestler, M.~Abello, M.~Roethke, F.~Sahm, W.~Wick,
  A.~M. Nagel, S.~Heiland, H.-P. Schlemmer, M.~Bendszus, J.~R. Reichenbach, and
  A.~Radbruch, ``Quantitative susceptibility mapping differentiates between
  blood depositions and calcifications in patients with glioblastoma,'' {\em
  PLOS ONE}, vol.~8, pp.~1--8, 03 2013.

\bibitem{Chen:QSM_calcification:2014}
W.~Chen, W.~Zhu, I.~Kovanlikaya, A.~Kovanlikaya, T.~Liu, S.~Wang, C.~Salustri,
  and Y.~Wang, ``Intracranial calcifications and hemorrhages: Characterization
  with quantitative susceptibility mapping,'' {\em Radiology}, vol.~270, no.~2,
  pp.~496--505, 2014.

\bibitem{Pruessmann:SENSE:1999}
K.~P. Pruessmann, M.~Weiger, M.~B. Scheidegger, and P.~Boesiger, ``Sense:
  Sensitivity encoding for fast mri,'' {\em Magnetic Resonance in Medicine},
  vol.~42, no.~5, pp.~952--962, 1999.

\bibitem{Griswold:GRAPPA:2002}
M.~A. Griswold, P.~M. Jakob, R.~M. Heidemann, M.~Nittka, V.~Jellus, J.~Wang,
  B.~Kiefer, and A.~Haase, ``Generalized autocalibrating partially parallel
  acquisitions (grappa),'' {\em Magnetic Resonance in Medicine}, vol.~47,
  no.~6, pp.~1202--1210, 2002.

\bibitem{Uecker:ESPIRiT:2014}
M.~Uecker, P.~Lai, M.~J. Murphy, P.~Virtue, M.~Elad, J.~M. Pauly, S.~S.
  Vasanawala, and M.~Lustig, ``Espirit—an eigenvalue approach to
  autocalibrating parallel mri: Where sense meets grappa,'' {\em Magnetic
  Resonance in Medicine}, vol.~71, no.~3, pp.~990--1001, 2014.

\bibitem{RUP06}
E.~J. Cand{\`e}s, J.~Romberg, and T.~Tao, ``Robust uncertainty principles:
  Exact signal reconstruction from highly incomplete frequency information,''
  {\em IEEE Trans. Inf. Theory}, vol.~52(2), pp.~489--509, 2006.

\bibitem{CS06}
D.~L. Donoho, ``Compressed sensing,'' {\em IEEE Trans. Inf. Theory}, vol.~52,
  no.~4, pp.~1289--1306, 2006.

\bibitem{Block:ModelT2:2009}
K.~T. {Block}, M.~{Uecker}, and J.~{Frahm}, ``Model-based iterative
  reconstruction for radial fast spin-echo mri,'' {\em IEEE Transactions on
  Medical Imaging}, vol.~28, no.~11, pp.~1759--1769, 2009.

\bibitem{Zhao:MR_mapping:2014}
B.~{Zhao}, F.~{Lam}, and Z.~{Liang}, ``Model-based mr parameter mapping with
  sparsity constraints: Parameter estimation and performance bounds,'' {\em
  IEEE Transactions on Medical Imaging}, vol.~33, no.~9, pp.~1832--1844, 2014.

\bibitem{Tamir:T2:2017}
J.~I. Tamir, M.~Uecker, W.~Chen, P.~Lai, M.~T. Alley, S.~S. Vasanawala, and
  M.~Lustig, ``T2 shuffling: Sharp, multicontrast, volumetric fast spin-echo
  imaging,'' {\em Magnetic Resonance in Medicine}, vol.~77, no.~1,
  pp.~180--195, 2017.

\bibitem{l1stable06}
E.~J. Cand{\`e}s, J.~K. Romberg, and T.~Tao, ``Stable signal recovery from
  incomplete and inaccurate measurements,'' {\em Communications on Pure and
  Applied Mathematics}, vol.~59, no.~8, pp.~1207--1223, 2006.

\bibitem{Yang2010ARO}
A.~Y. Yang, A.~Ganesh, Z.~Zhou, S.~S. Sastry, and Y.~Ma, ``A review of fast
  l1-minimization algorithms for robust face recognition,'' {\em CoRR},
  vol.~abs/1007.3753, 2010.

\bibitem{Tetko:Overfitting:1995}
I.~V. Tetko, D.~J. Livingstone, and A.~I. Luik, ``Neural network studies. 1.
  comparison of overfitting and overtraining,'' {\em Journal of Chemical
  Information and Computer Sciences}, vol.~35, pp.~826--833, 1995.

\bibitem{Hawkins:Overfitting:2004}
D.~M. Hawkins, ``The problem of overfitting,'' {\em Journal of Chemical
  Information and Computer Sciences}, vol.~44, pp.~1--12, 2004.

\bibitem{Hansen:l_curve:2000}
P.~C. Hansen, ``The l-curve and its use in the numerical treatment of inverse
  problems,'' in {\em in Computational Inverse Problems in Electrocardiology,
  ed. P. Johnston, Advances in Computational Bioengineering}, pp.~119--142, WIT
  Press, 2000.

\bibitem{Srivastava:2016:wave_denoise}
M.~Srivastava, C.~L. Anderson, and J.~H. Freed, ``A new wavelet denoising
  method for selecting decomposition levels and noise thresholds,'' {\em IEEE
  Access}, vol.~4, pp.~3862--3877, 2016.

\bibitem{Khare:MRM:2012}
K.~Khare, C.~J. Hardy, K.~F. King, P.~A. Turski, and L.~Marinelli,
  ``Accelerated mr imaging using compressive sensing with no free parameters,''
  {\em Magnetic Resonance in Medicine}, vol.~68, no.~5, pp.~1450--1457, 2012.

\bibitem{Ahmad:TCI:2015}
R.~Ahmad and P.~Schniter, ``Iteratively reweighted approaches to sparse
  composite regularization,'' {\em IEEE Transactions on Computational Imaging},
  vol.~1, no.~4, pp.~220--235, 2015.

\bibitem{Rangan:GAMP:2011}
S.~Rangan, ``Generalized approximate message passing for estimation with random
  linear mixing,'' in {\em Proceedings of IEEE ISIT}, pp.~2168--2172, July
  2011.

\bibitem{PE_GAMP17}
S.~Huang and T.~D. Tran, ``Sparse signal recovery using generalized approximate
  message passing with built-in parameter estimation,'' in {\em Proceedings of
  IEEE ICASSP}, pp.~4321--4325, March 2017.

\bibitem{Donoho:AMP:2009}
D.~L. Donoho, A.~Maleki, and A.~Montanari, ``Message-passing algorithms for
  compressed sensing,'' {\em Proceedings of the National Academy of Sciences},
  vol.~106, no.~45, pp.~18914--18919, 2009.

\bibitem{Baron:2010}
D.~{Baron}, S.~{Sarvotham}, and R.~G. {Baraniuk}, ``Bayesian compressive
  sensing via belief propagation,'' {\em IEEE Trans. Signal Process.}, vol.~58,
  no.~1, pp.~269--280, 2010.

\bibitem{Guo:SURE:2015}
C.~{Guo} and M.~E. {Davies}, ``Near optimal compressed sensing without priors:
  Parametric sure approximate message passing,'' {\em IEEE Trans. Signal
  Process.}, vol.~63, no.~8, pp.~2130--2141, 2015.

\bibitem{Metzler:Denoising:2016}
C.~A. {Metzler}, A.~{Maleki}, and R.~G. {Baraniuk}, ``From denoising to
  compressed sensing,'' {\em IEEE Trans. Inf. Theory}, vol.~62, pp.~5117--5144,
  Sep. 2016.

\bibitem{Ma:AMP_Denoise:2016}
Y.~{Ma}, J.~{Zhu}, and D.~{Baron}, ``Approximate message passing algorithm with
  universal denoising and gaussian mixture learning,'' {\em IEEE Trans. on
  Signal Process.}, vol.~64, no.~21, pp.~5611--5622, 2016.

\bibitem{Krzakala:2012:1}
F.~Krzakala, M.~M\'ezard, F.~Sausset, Y.~F. Sun, and L.~Zdeborov\'a,
  ``Statistical-physics-based reconstruction in compressed sensing,'' {\em
  Phys. Rev. X}, vol.~2, p.~021005, May 2012.

\bibitem{Vila:EMGM:2013}
J.~P. {Vila} and P.~{Schniter}, ``Expectation-maximization gaussian-mixture
  approximate message passing,'' {\em IEEE Trans. Signal Process.}, vol.~61,
  no.~19, pp.~4658--4672, 2013.

\bibitem{Kamilov:PE:2014}
U.~S. Kamilov, S.~Rangan, A.~K. Fletcher, and M.~Unser, ``Approximate message
  passing with consistent parameter estimation and applications to sparse
  learning,'' {\em IEEE Trans. Inf. Theory}, vol.~60, pp.~2969--2985, May 2014.

\bibitem{Krzakala:2012:2}
F.~Krzakala, M.~M{\'{e}}zard, F.~Sausset, Y.~Sun, and L.~Zdeborov{\'{a}},
  ``Probabilistic reconstruction in compressed sensing: algorithms, phase
  diagrams, and threshold achieving matrices,'' {\em J. Stat. Mech. Theory
  Exp.}, vol.~2012, p.~P08009, aug 2012.

\bibitem{Ziniel:DCS:2013}
J.~{Ziniel} and P.~{Schniter}, ``Dynamic compressive sensing of time-varying
  signals via approximate message passing,'' {\em IEEE Transactions on Signal
  Processing}, vol.~61, no.~21, pp.~5270--5284, 2013.

\bibitem{Millard:AMP_MRI:2020}
C.~{Millard}, A.~T. {Hess}, B.~{Mailhe}, and J.~{Tanner}, ``An approximate
  message passing algorithm for rapid parameter-free compressed sensing mri,''
  in {\em 2020 IEEE International Conference on Image Processing (ICIP)},
  pp.~91--95, 2020.

\bibitem{Qiao:AMP_MRI:2020}
X.~{Qiao}, J.~{Du}, L.~{Wang}, Z.~{He}, and Y.~{Jia}, ``A model-based deep
  network for mri reconstruction using approximate message passing algorithm,''
  in {\em 2020 IEEE International Conference on Acoustics, Speech and Signal
  Processing (ICASSP)}, pp.~1105--1109, 2020.

\bibitem{Rich:AMP_PC_MRI:2016}
A.~Rich, L.~C. Potter, N.~Jin, J.~Ash, O.~P. Simonetti, and R.~Ahmad, ``A
  bayesian model for highly accelerated phase-contrast mri,'' {\em Magnetic
  Resonance in Medicine}, vol.~76, no.~2, pp.~689--701, 2016.

\bibitem{Rich:4Dflow:2018}
A.~Rich, L.~C. Potter, N.~Jin, Y.~Liu, O.~P. Simonetti, and R.~Ahmad, ``A
  bayesian approach for 4d flow imaging of aortic valve in a single
  breath-hold,'' {\em Magnetic Resonance in Medicine}, vol.~81, no.~2,
  pp.~811--824, 2019.

\bibitem{Aaron:4Dflow:2020}
A.~Pruitt, A.~Rich, Y.~Liu, N.~Jin, L.~Potter, M.~Tong, S.~Rajpal,
  O.~Simonetti, and R.~Ahmad, ``Fully self-gated whole-heart 4d flow imaging
  from a 5-minute scan,'' {\em Magnetic Resonance in Medicine}, vol.~85, no.~3,
  pp.~1222--1236, 2020.

\bibitem{Liu:MEDI:2013}
T.~Liu, C.~Wisnieff, M.~Lou, W.~Chen, P.~Spincemaille, and Y.~Wang, ``Nonlinear
  formulation of the magnetic field to source relationship for robust
  quantitative susceptibility mapping,'' {\em Magnetic Resonance in Medicine},
  vol.~69, no.~2, pp.~467--476, 2013.

\bibitem{Liu:MEDI:2012}
J.~Liu, T.~Liu, L.~{de Rochefort}, J.~Ledoux, I.~Khalidov, W.~Chen, A.~J.
  Tsiouris, C.~Wisnieff, P.~Spincemaille, M.~R. Prince, and Y.~Wang,
  ``Morphology enabled dipole inversion for quantitative susceptibility mapping
  using structural consistency between the magnitude image and the
  susceptibility map,'' {\em NeuroImage}, vol.~59, no.~3, pp.~2560--2568, 2012.

\bibitem{MRI:Nishimura:2010}
D.~G. Nishimura, {\em Principles of Magnetic Resonance Imaging}.
\newblock Stanford, CA, USA: Stanford University, 2010.

\bibitem{DBWav92}
I.~Daubechies, {\em Ten lectures on wavelets}.
\newblock Philadelphia, PA, USA: Society for Industrial and Applied
  Mathematics, 1992.

\bibitem{Bayati:SE:2011}
M.~{Bayati} and A.~{Montanari}, ``The dynamics of message passing on dense
  graphs, with applications to compressed sensing,'' {\em IEEE Trans. Inf.
  Theory}, vol.~57, no.~2, pp.~764--785, 2011.

\bibitem{Rangan:DampingCvg:2014}
S.~{Rangan}, P.~{Schniter}, and A.~{Fletcher}, ``On the convergence of
  approximate message passing with arbitrary matrices,'' in {\em Proceedings of
  IEEE ISIT}, pp.~236--240, 2014.

\bibitem{Vila:DampingMR:2015}
J.~{Vila}, P.~{Schniter}, S.~{Rangan}, F.~{Krzakala}, and L.~{Zdeborov\'a},
  ``Adaptive damping and mean removal for the generalized approximate message
  passing algorithm,'' in {\em Proceedings of IEEE ICASSP}, pp.~2021--2025,
  2015.

\bibitem{Kschischang:2001}
F.~R. {Kschischang}, B.~J. {Frey}, and H.~A. {Loeliger}, ``Factor graphs and
  the sum-product algorithm,'' {\em IEEE Trans. Inf. Theory}, vol.~47, no.~2,
  pp.~498--519, 2001.

\bibitem{Koller:2009}
D.~Koller and N.~Friedman, {\em Probabilistic Graphical Models: Principles and
  Techniques - Adaptive Computation and Machine Learning}.
\newblock The MIT Press, 2009.

\bibitem{Minka:2001}
T.~P. Minka, {\em A Family of Algorithms for Approximate Bayesian Inference}.
\newblock PhD thesis, Massachusetts Institute of Technology, USA, Jan. 2001.

\bibitem{Minka:Div:2005}
T.~Minka, ``Divergence measures and message passing,'' Tech. Rep.
  MSR-TR-2005-173, Microsoft Research Ltd., Cambridge, UK, January 2005.

\bibitem{WITTEN2011147}
I.~H. Witten, E.~Frank, and M.~A. Hall, ``Chapter 5 - credibility: Evaluating
  what's been learned,'' in {\em Data Mining: Practical Machine Learning Tools
  and Techniques}, pp.~147--187, Boston: Morgan Kaufmann, third edition~ed.,
  2011.

\bibitem{Beck:FISTA:2009}
A.~Beck and M.~Teboulle, ``A fast iterative shrinkage-thresholding algorithm
  for linear inverse problems,'' {\em SIAM Journal on Imaging Sciences},
  vol.~2, no.~1, pp.~183--202, 2009.

\bibitem{Laplacian:L1:2014}
W.~Li, A.~V. Avram, B.~Wu, X.~Xiao, and C.~Liu, ``Integrated laplacian-based
  phase unwrapping and background phase removal for quantitative susceptibility
  mapping,'' {\em NMR in Biomedicine}, vol.~27, no.~2, pp.~219--227, 2014.

\bibitem{PDF:Liu:2011}
T.~Liu, I.~Khalidov, L.~de~Rochefort, P.~Spincemaille, J.~Liu, A.~J. Tsiouris,
  and Y.~Wang, ``A novel background field removal method for mri using
  projection onto dipole fields (pdf),'' {\em NMR in Biomedicine}, vol.~24,
  no.~9, pp.~1129--1136, 2011.

\end{thebibliography}

\newpage

{\bfseries\Large Supporting Information}

Additional Supporting Information may be found online in the Supporting Information section.

\paragraph{Supporting Figure S1} Retrospective undersampling: sagittal views of recovered QSM using the least squares approach (LSQ), the $l_1$-norm regularization approach with parameter tuning (L1-T) and L-curve method (L1-L), the proposed AMP-PE approach.

\paragraph{Supporting Figure S2} Retrospective undersampling: coronal views of recovered QSM using the least squares approach (LSQ), the $l_1$-norm regularization approach with parameter tuning (L1-T) and L-curve method (L1-L), the proposed AMP-PE approach.

\paragraph{Supporting Figure S3} Retrospective undersampling: axial views of recovered local field maps using the least squares approach (LSQ), the $l_1$-norm regularization approach with parameter tuning (L1-T) and L-curve method (L1-L), the proposed AMP-PE approach.

\paragraph{Supporting Figure S4} Retrospective undersampling: sagittal views of recovered local field maps using the least squares approach (LSQ), the $l_1$-norm regularization approach with parameter tuning (L1-T) and L-curve method (L1-L), the proposed AMP-PE approach.

\paragraph{Supporting Figure S5} Retrospective undersampling: coronal views of recovered local field maps using the least squares approach (LSQ), the $l_1$-norm regularization approach with parameter tuning (L1-T) and L-curve method (L1-L), the proposed AMP-PE approach.

\paragraph{Supporting Figure S6} Prospective undersampling: recovered initial magnetization $\widehat{\vz}_0$ using the least squares approach (LSQ), the $l_1$-norm regularization approach with the L-curve method (L1-L), the proposed AMP-PE approach.

\paragraph{Supporting Figure S7} Prospective undersampling: recovered $R_2^*$ map $\widehat{\vr}_2^*$ using the least squares approach (LSQ), the $l_1$-norm regularization approach with the L-curve method (L1-L), the proposed AMP-PE approach.

\paragraph{Supporting Figure S8} Prospective undersampling: axial view of recovered QSM using the least squares approach (LSQ), the $l_1$-norm regularization approach with the L-curve method (L1-L), the proposed AMP-PE approach.

\paragraph{Supporting Figure S9} Prospective undersampling: sagittal view of recovered QSM using the least squares approach (LSQ), the $l_1$-norm regularization approach with the L-curve method (L1-L), the proposed AMP-PE approach.

\paragraph{Supporting Figure S10} Prospective undersampling: coronal view of recovered QSM using the least squares approach (LSQ), the $l_1$-norm regularization approach with the L-curve method (L1-L), the proposed AMP-PE approach.

\paragraph{Supporting Figure S11} The Poisson-disk sampling pattern produces a more uniform sampling across k-space than the random sampling pattern.

\paragraph{Supporting Figure S12} Comparison of the recovered initial magnetizations $\hat{\vz}_0$ using random sampling and Poisson-disk sampling with the proposed AMP-PE approach. The Poisson-disk sampling pattern leads to lower errors.

\paragraph{Supporting Figure S13} Comparison of the recovered $R_2^*$ map $\hat{\vr}_2^*$ using random sampling and Poisson-disk sampling with the proposed AMP-PE approach. The Poisson-disk sampling pattern leads to lower errors.

\paragraph{Supporting Figure S14} Comparison of the recovered QSM using random sampling and Poisson-disk sampling with the proposed AMP-PE approach. The Poisson-disk sampling pattern leads to lower errors.

\paragraph{Supporting Figure S15} Comparison of the recovered $\hat{\vz}_1$ using Bernoulli-Gaussian-mixture prior and the Laplace prior with the proposed AMP-PE approach. The Laplace prior leads to lower errors.

\paragraph{Supporting Table S1} Retrospective undersampling (P1-R): pixel-wise absolute errors of recovered images across different subjects..

\paragraph{Supporting Table S2} Retrospective undersampling (P2-R): pixel-wise absolute errors of recovered images across different subjects.

\paragraph{Supporting Table S3} Retrospective undersampling: HFEN values of recovered QSM $\hat{\chi}$.

\paragraph{Supporting Table S4} Retrospective undersampling: normalized absolute errors of recovered local fields.

\paragraph{Supporting Table S5} Retrospective undersampling: pixel-wise absolute errors of recovered local fields across different subjects.

\paragraph{Supporting Table S6} Prospective undersampling (P1-P): normalized absolute errors of recovered images.

\paragraph{Supporting Table S7} Prospective undersampling (P2-P): normalized absolute errors of recovered images.

\paragraph{Supporting Table S8} Prospective undersampling (P1-P): pixel-wise absolute errors of recovered images across different subjects.

\paragraph{Supporting Table S9} Prospective undersampling (P2-P): pixel-wise absolute errors of recovered images across different subjects.

\paragraph{Supporting Table S10} Parameters in the $l_1$-norm regularization approach. For retrospective undersampling, the 1st (S1) and 8th (S8) subjects are used as training data, the rest are used as test data.

\paragraph{Supporting Table S11} Retrospective undersampling (P1-R): normalized absolute errors of recovered images from L1 with Exhaustive search (L1-E) and AMP.

\paragraph{Supporting Table S12} Retrospective undersampling (P2-R): normalized absolute errors of recovered images from L1 with Exhaustive search (L1-E) and AMP.

\paragraph{Supporting Algorithm S1} Recovery of the multi-echo image distribution $p_{\mathcal{M}}(z_{in}|\vy)$.

\paragraph{Supporting Algorithm S2} Recovery of $R_2^*$ map $\vr_2^*$, initial magnetization $\vz_0$ and multi-echo image $\vz_i$.

\end{document}